%% file: main.tex
\documentclass{PoS}
\usepackage{epsfig,dcolumn}
\usepackage{color}
\usepackage{amsmath}
\usepackage{amsfonts}
\usepackage{amssymb}
\usepackage[T1]{fontenc}       
\usepackage[utf8]{inputenc}
\usepackage[dvipsnames]{xcolor}
\usepackage{multirow}

\usepackage{graphicx}
\usepackage{type1cm}
\usepackage{eso-pic}
\usepackage{tabularx}

\newcommand{\be}{\begin{equation}}
\newcommand{\ee}{\end{equation}}

\newcommand{\im}{\mathrm{Im}\,}
\newcommand{\re}{\mathrm{Re}\,}

\usepackage{xspace}

\newcommand{\gevnospace}{\ensuremath{{\mathrm{\,Ge\kern -0.1em V}}}}

\newcommand{\mevnospace}{\ensuremath{{\mathrm{\,Me\kern -0.1em V}}}}

\graphicspath{{figures/}}

\usepackage{bm} 

\title{Dispersive determination of low energy $\pi K$ interactions}

\ShortTitle{Low energy $\pi K$ interactions}

\author{\speaker{A.~Rodas}\\
Department of Physics, College of William and Mary, Williamsburg, VA 23187, USA \\
Theory Center, Thomas  Jefferson  National  Accelerator  Facility, 
Newport  News,  VA  23606,  USA \\
        E-mail: \email{arodas@wm.edu}}

\author{J.R.~Pel\'aez \\
  Departamento de F\'isica Te\'orica and IPARCOS, Universidad Complutense de Madrid, 28040 Madrid, Spain\\
        E-mail: \email{jrpelaez@fis.ucm.es}}

\abstract{We summarize the application of dispersion relations for the determination of low energy $\pi K$ interactions. In particular, we present our recent dispersive analyses for the complete study of both $\pi K\to \pi K$ and $\pi \pi \rightarrow K \bar K$ channels. First, we summarize our approach, we study both channels by using several families of dispersion relations and present the most relevant partial waves. Then we apply the dispersive formalism to the threshold region, where we extract many different low-energy parameters.  }

\FullConference{%
  The 10th International Workshop on Chiral Dynamics - CD2021\\
  15-19 November 2021\\
  Online
  }

\begin{document}

\section{Introduction}

Even though the formulation of quantum Chromodynamics (QCD) is known since half a century ago, the first-principles determination of its low energy interactions remains a mystery. The strong interaction grows as the energies involved become smaller. This prevents us from using perturbation theory techniques. Actually, at low energies, the chiral symmetry is spontaneously broken. Quarks are confined into hadrons at the 1 GeV scale. As a result, the well-known pseudo-Goldstone boson octet emerges, of which pions and kaons are the two lightest particles. Not surprisingly, the study of these meson-meson interactions is crucial for testing our understanding of the QCD chiral symmetry breaking pattern. Furthermore, light mesons appear in the final states of most hadronic interactions in experiments, which makes the understanding of their scattering processes necessary. On top of that, there exists a myriad of mesons that appear as resonances in these low-energy meson-meson scattering channels. For example, this is the case of the $\sigma/f_0(500)$ and $\kappa/K^*_0(700)$ resonances. These have been subject to fierce debate for many decades~\cite{Pelaez:2015qba}, they do not fit in the ordinary $q \bar q$ picture, and could probably exhibit more exotic structures~\cite{Esposito:2016noz,Guo:2017jvc}.

Unfortunately, mesons are short-lived. They decay through electroweak processes. This renders their direct experimental detection very challenging, as no two-meson beam experiment could be realistically run. We rely on indirect measurements instead, of which there exist several $\pi K$ scattering experiments  \cite{Bakker:1970wg, Cho:1969qk,Linglin:1973ci, Estabrooks:1977xe, Aston:1987ir} and the recent KLF proposal~\cite{KLF:2020gai}. All the main data sets are obtained by studying meson-nucleon to meson-meson-nucleon interactions. The desired pion-meson scattering is then obtained by assuming one pion exchange dominance, which in practice introduces a source of systematic uncertainty. This could produce a large systematic effect depending on the precision and energy range for the determination of the transferred momentum. As a result, the literature is plagued with many conflicting data sets (see \cite{Pelaez:2015qba, Pelaez:2020gnd, Pelaez:2021dak}). Unfortunately, some of the $\pi K$ scattering measurements seem to be largely affected by these.

The determination of their low-energy interactions is crucial to testing our knowledge of QCD in the non-perturbative regime. The effective field theory describing meson interactions, Chiral Perturbation Theory (ChPT), is known for many decades now. Pions, kaons, and etas form what is known as the $SU(3)$ pseudoscalar meson octet. As a result, the scattering processes can be described at a given perturbative order by making use of phenomenological lagrangians~\cite{Weinberg:1978kz,Gasser:1983yg,Bernard:1990kw}, fulfilling the basic symmetries of QCD. Then, the free parameters obtained can be fixed if one has enough experimental data~\cite{Knecht:1995tr,Bijnens:1995yn,Bijnens:1999sh}. 

However, the perturbative nature of the ChPT expansion violates unitarity, which is a basic principle of our amplitudes. This problem is aggravated when dealing with cross-section enhancements, which is the situation involving resonances. In order to circumvent this problem, several unitarization schemes were developed in the past~\cite{Dobado:1996ps,Oller:1997ng,Oller:1998hw,GomezNicola:2001as}. This methodology has been successful when describing parts of the meson sector. Unfortunately, it is not fully model-independent, and it is not well suited for high accuracy.

Of all the resonances appearing in $\pi K$ scattering, the $\kappa/K^*_0(700)$ is the most interesting one. It appears close to the scalar $\pi K$ threshold, hinting at a possible molecular nature, and it is too broad to be determined with precision using simple approaches. In order to extract it with robustness, one needs to rely on suitable techniques for performing a stable analytic continuation far from the real axis, where the poles associated with such resonances are expected. Note that this problem is shared by lattice QCD determinations~\cite{Briceno:2016mjc,Wilson:2019wfr,Rendon:2020rtw}.

For all the reasons mentioned above, making use of the $S$-matrix principles has become an interesting tool in modern amplitude analyses. When constraining our amplitudes by imposing unitarity, analyticity, and crossing symmetry we obtain a system of integral equations, called dispersion relations. These allow us first to determine the systematic effects present in the data and parameterizations. They can later be used to eliminate these unwanted effects if correctly imposed over data. Because they are built through the Cauchy Theorem, these integral equations perform a very stable continuation to the complex plane. Hence they enable us to robustly extract resonance parameters. Actually, the low energy parameters can also be determined with precision up to high orders of the expansion.

In this proceeding, we briefly summarize previous works~\cite{Pelaez:2020gnd,Pelaez:2016tgi,Pelaez:2016klv,Pelaez:2018qny,Pelaez:2020uiw} where the $\pi K\to \pi K$ and $\pi \pi \to K \bar K$ interactions are analyzed by means of a dispersive formulation. We will first introduce the main tools and basic concepts, in order to perform the analysis. We then explain how the constraints over experimental data are performed and present some of the final parameterizations. Finally, we provide the reader with an abundant amount of low-energy parameters and compare them with previous analyses.

\section{Dispersive analysis}

 Let us briefly recall the derivation of a dispersion relation. Starting from a meson-meson amplitude $F(s,t,u)$, which depends just on two Mandelstam variables, since $s+t+u=\sum_i m_i^2$. Dispersion relations are nothing but the implementation of the $S$-matrix principles of analyticity, a consequence of causality, and crossing symmetry. In particular, they make use of the Mandelstam Hypothesis, which estates the analyticity of these amplitudes in the complex $s,t$ plane. Of all the possible ways of building a dispersion relation, the simplest one is obtained by fixing the $t$ variable. It yields two main singularity structures, a right-hand cut (RHC) coming from unitarity in the physical region and a left-hand cut (LHC), created by crossed channel interactions. Once the analytic structure is known, one can invoke the Cauchy theorem and produce a dispersion relation 
\begin{equation}
    F(s,t)=\frac{1}{\pi}\int_{s_{th}}^{\infty}ds' \frac{\im F(s',t)}{s'-s}+\frac{1}{\pi}\int_{u_{th}}^{-\infty}ds' \frac{\im F(s',t)}{s'-s},
\end{equation}
where the first and second integrals correspond to the RHC and LHC, respectively. For example, the cross channel contribtuion to $\pi K$ scattering is $\pi \pi \to K\bar K$. It is worth noting that subtractions may be needed depending on the behavior of the amplitude at  high-energy. 

 If one has to determine each one of the relevant partial waves, then the expansion and posterior projection of all the amplitudes into partial waves inside the dispersion relations is required. For $\pi K$ interactions it means that we must obtain the description of both channels ($\pi K \to \pi K$ and $\pi \pi \rightarrow K \bar K$) at the same time. If utmost accuracy is not the main goal, one could get rid of part of these LHC contributions to perform a simpler analysis. One could for example make use of the Inverse Amplitude Method~\cite{Dobado:1996ps,GomezNicola:2007qj}, which correctly implements the RHC physics into a dispersive formulation. There are also some works where the unphysical LHC cuts have been approximated, but the unitarity cut is described from the data using dispersion relations \cite{Zheng:2003rw,Yao:2020bxx, Danilkin:2020pak}.

Perhaps one of the most salient features of the dispersion relations is that they produce a stable extrapolation to the complex plane. In practice, this means that we can achieve both a precise and robust determination of the $\kappa/K^*_0(700)$ pole position.

The simplest dispersive set is called Forward Dispersion Relations. They are obtained by setting $t=0$ for the amplitude $F(s,t)$. Apart from the main advantage of its simplicity, they can also be applied up to arbitrarily high energies, thus constraining the amplitudes in the inelastic region. Such a set was proven to be very useful in previous analyses \cite{Pelaez:2016tgi,Pelaez:2004vs}.

As for the implementation of partial wave dispersion relations, the best known are the fixed-$t$ Roy-Steiner dispersion relations \cite{Roy:1971tc,Steiner:1971ms}, . Which, in the particular case of $\pi K$ scattering read
\begin{equation}
    f^I_l(s)=\frac{m_+a^+_0}{2}+\frac{1}{\pi}\sum_\ell \int^{\infty}_{m_+^2}ds' L^I_{l, \ell}(s,s') \im f^I_\ell(s')
+\frac{1}{\pi}\sum_{\ell\geq0}\int^{\infty}_{4m_{\pi}^2} dt' L^0_{l, 2\ell}(s,t') \im g^0_{2\ell}(t'),
\end{equation}
where $f^I_\ell,\, g^I_\ell$ stand for the $\pi K\to \pi K$ and $\pi\pi \to K \bar K$ partial waves, respectively.

 For $\pi K$ interactions, fixed-$t$ dispersion relations cannot be applied far from the real axis, and they do not reach the physical regions of the $\pi \pi \to K \bar K$ channel. A different set, called hyperbolic dispersion relations~\cite{Hite:1973pm} must be used. They are defined through the hyperbolic relation $(s-a)(u-a)=b$, where $a$ can be chosen to optimize the region where the dispersion relations can be applied, both in the real axis and in the complex plane. Both of these sets have produced sound determinations in the recent past ( for $\pi \pi$ \cite{Ananthanarayan:2000ht,Colangelo:2001df,Kaminski:2002pe,GarciaMartin:2011cn,Kaminski:2011vj,Moussallam:2011zg,Caprini:2011ky,Albaladejo:2018gif,Pelaez:2019eqa}, for $\pi N$ \cite{Ditsche:2012fv, Hoferichter:2015hva}, for $e^+ e^- \to \pi^+ \pi^-$ \cite{Colangelo:2018mtw}, for  $\gamma^{(*)} \gamma^{(*)} \to \pi \pi$ \cite{Hoferichter:2011wk,Moussallam:2013una, Danilkin:2018qfn, Hoferichter:2019nlq},
and for $\pi K$ \cite{Johannesson:1976qp,Ananthanarayan:2000cp,Ananthanarayan:2001uy,Buettiker:2003pp,DescotesGenon:2006uk}).

There are two different approaches towards implementing dispersion relations in the literature. The first one solves the dispersion relations below a matching point $s_m$~\cite{Ananthanarayan:2000ht,Colangelo:2001df,Buettiker:2003pp}, above which the partial waves are just mere input coming from a fit to the data, together with theoretical input for the low energy parameters. In this sense, one does not include data close to threshold. The solution is a prediction coming from the dispersion relations. The second approach is to constrain the initial fits to the data together with the dispersion relations~\cite{GarciaMartin:2011cn,Kaminski:2011vj,Pelaez:2020gnd}. The result is a set of fits that describe both the data and the $S$-matrix requirements. The number of subtractions will depend on the asymptotic region, and on the preferred approach.

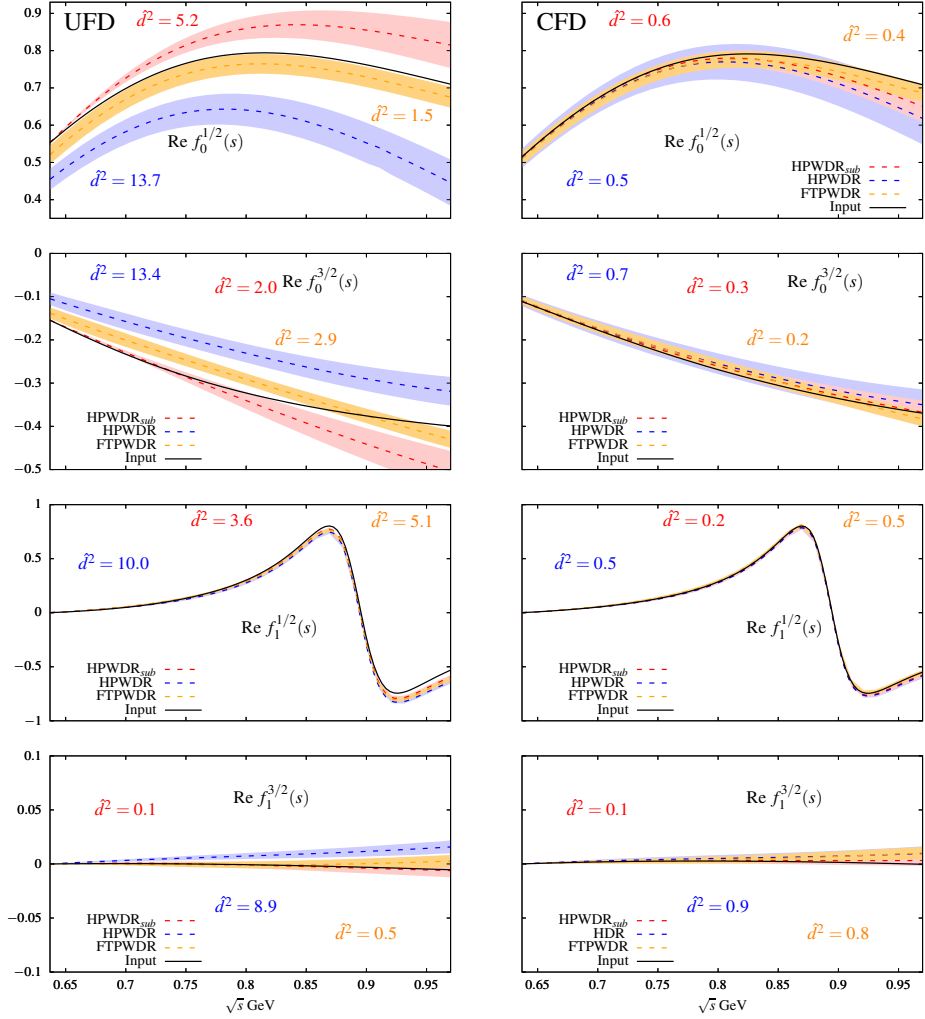
\begin{figure}[!ht]
\centering
\resizebox{0.8\textwidth}{!}{\input{figures/kappa12ufd.tex} \hspace{0.5cm} \input{figures/kappa12cfd.tex}}\\
\resizebox{0.8\textwidth}{!}{\input{figures/kappa32ufd.tex} \hspace{0.5cm} \input{figures/kappa32cfd.tex}}\\
\resizebox{0.8\textwidth}{!}{\input{figures/kappa121ufd.tex} \hspace{0.5cm} \input{figures/kappa121cfd.tex}}\\
\resizebox{0.8\textwidth}{!}{\input{figures/kappa321ufd.tex} \hspace{0.5cm} \input{figures/kappa321cfd.tex}}\\
\caption{Dispersive and direct results for UFD (left) versus CFD (right) real parts of the $f^{1/2}_0(s),f^{3/2}_0(s),f^{1/2}_1(s)$ and $f^{3/2}_1(s)$ partial waves. The uncertainty bands correspond to the uncertainty in the difference between the input and the respective dispersive representation. Each color corresponds to a different dispersion relation.}
\label{fig:alltogether}
\end{figure}

In our analyses on the $\pi K$ system, we found that combining both hyperbolic (HPWDR) and fixed-$t$ (FTPWDR) dispersion relations with different numbers of substractions was beneficial. The fixed-$t$ are only used for analyzing $\pi K \to \pi K$ on the real axis, but the hyperbolic ones are applied for both $\pi K \to \pi K$ and $\pi \pi \to K \bar K$. They are also used to extract the resonances on the complex plane. On top of that, we will include Forward Dispersion Relations to constrain the data at higher energies, including the contributions arising from the Regge exchanges. To summarize, we are making use of 16 dispersion relations and 13 partial waves. We refer the reader to~\cite{Pelaez:2020gnd} for more details.

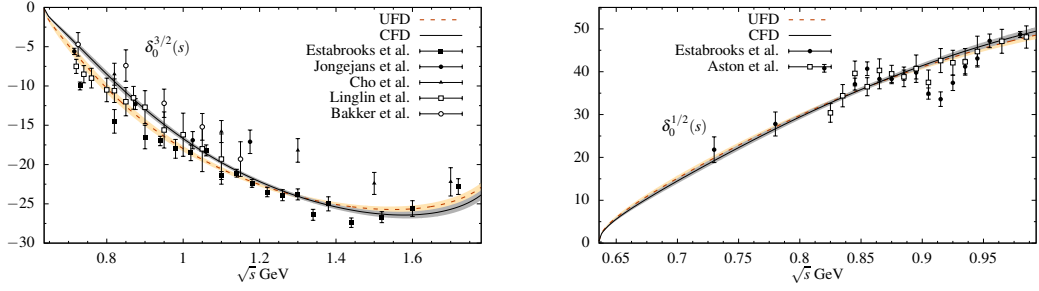
\begin{figure}
\resizebox{0.45\textwidth}{!}{\input{figures/selas32cfd.tex}} \hspace{0.3cm} \resizebox{0.45\textwidth}{!}{\input{figures/selascfd.tex}}
\caption{$f^{3/2}_0(s)$ (left) and $f^{1/2}_0(s)$ (right) CFD vs UFD partial waves.}
\label{fig:cfd}
\end{figure}

The first stage is to obtain a simple yet flexible description of the partial waves from the data sets. To do so, we combine the implementation of different functional forms, together with a consistent way of pruning the data \cite{NavarroPerez:2015gaz}. The result is a complete description of all relevant partial waves in both channels up to roughly 1.8-2 GeV. Thereupon the dispersion relations are applied and the consistency between the fit to the data and the dispersive output is tested. In order to ``measure'' the deviation, we define the following penalty function
\begin{equation} 
d^2=\frac{1}{N}\sum_{i=1}^{N} \left(\frac{d_i}{\Delta d_i,}\right)^2, 
\end{equation}
where $d_i$ is
the difference between the ``input'' and ``output'', weighted by the relative uncertainty between the two. These are evaluated at a uniform grid in the energy $\sqrt{s_i}$. This ``distance'' is now included in the original $\chi^2$ to the data, and they are weighted by the degrees of freedom they are roughly describing. As shown in Fig.\ref{fig:alltogether} there is a clear deviation when using the original fits to the data (UFD). They exhibit systematic effects both in the data and the parameterizations.  The last step is to constrain these fits using dispersion relations. As shown in Figs.~\ref{fig:alltogether} and ~\ref{fig:cfd}. The final constrained result (CFD) is compatible with all dispersion relations and data. Actually, the deviations between the UFD and CFD fit, albeit small, are crucial to improve the dispersive description.

\section {Low energy parameters}

\begin{figure}[!hb]
    \centering
    \resizebox{0.8\textwidth}{!}{\colorbox{white}{\input{figures/sclcfd.tex}}}
    \caption{Comparison between various dispersive, lattice QCD, and ChPT determinations of the scalar scattering lengths. Our  ``Final value'' is shown as a blue ellipse and is listed in Table~\ref{tab:isospinscl}.}
    \label{fig:sclcfd}
\end{figure}
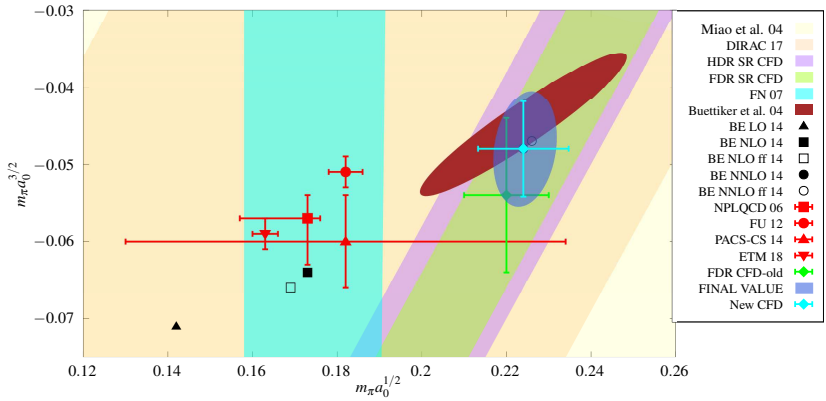

The dispersive formalism can also be applied to the threshold and sub-threshold regions. In this sense, one matches the partial wave expansion

\begin{equation}
    \frac{2}{\sqrt{s}}\re f^I_\ell(s)\simeq q^{2\ell}\left(a^I_\ell+b^I_\ell q^2+c^I_\ell q^4 +...\right),
    \label{eq:lowex}
\end{equation}
with the dispersive integral, by expanding the integrands in even powers of the center of mass momentum $q$. For example the scattering length for $F^-=\frac{1}{3}F^{1/2}-\frac{1}{3}F^{3/2}$ reads
\begin{equation}
a^-_0=\frac{ m_\pi m_K}{2 \pi^2 m_+}\!\!\int^{\infty}_{m_+^2}{\frac{\im F^-(s')}{(s'-m_-^ 2)(s'-m_+^2)}ds'}.
\label{eq:scla0fdr}
\end{equation}

At present, there is a great deal of interest in the values of the low energy parameters from ChPT, dispersion theory, and lattice QCD, since tension exists between the determinations using these different techniques. This is illustrated in Fig.~\ref{fig:sclcfd} when comparing the $S$-wave scattering lengths. We depict in the figure the $I=1/2,\,3/2$ values and include several references~\cite{Buettiker:2003pp,Miao:2004gy,Beane:2006gj,Flynn:2007ki,Fu:2011wc,Sasaki:2013vxa,Bijnens:2014lea,DIRAC:2014gmw,Pelaez:2016tgi,Helmes:2018nug} to be compared with our dispersive extraction in blue, listed in Table~\ref{tab:isospinscl}. It is worth noting that all our dispersive results, CFD fits and sum rules converge to the same region, showcasing the accuracy and robustness of our determination. 
\begin{center}
\begin{table}[h]
\centering
\footnotesize
\begin{tabular}{l|c|c|c|c|c}
\hline
&\multicolumn{3}{c|} {This work sum rules with CFD input}  & Sum rules \cite{Buettiker:2003pp}&  NNLO ChPT  \\
 & FDR & HDR & \textbf{Final Value}  &  Fixed-$t$ &  \cite{Bijnens:2004bu} and 
\cite{Bijnens:2014lea}$^*$\\
\hline
\hline
\rule[-0.175cm]{-0.1cm}{.5cm} $m_\pi a_0^{1/2}$ & 0.226$\pm$ 0.010 &  0.225$\pm$ 0.012 & \textbf{0.225$\pm$0.008}  & 0.224$\pm$0.022 & 0.224$^*$  \\
\rule[-0.175cm]{-0.1cm}{.5cm} $m_\pi a_0^{3/2}$  $\times$ 10&  $-$0.489$\pm$0.052 & $-$0.485$\pm$0.066 & \textbf{$-$0.480$\pm$0.067}& $-$0.448$\pm$0.077 & $-$0.471$^*$ \\
\hline
\end{tabular}
\caption{Determination of the $\pi K$ scalar scattering lengths using sum rules for $a_0^-$ with our CFD as input.}
\label{tab:isospinscl}
\end{table}
\end{center}

The scattering lengths are highly correlated to the position of the $S$-waves Adler Zeroes, which appear as a consequence of the QCD chiral symmetry breaking. We summarize in Table~\ref{tab:adlerzeros} our dispersive values for these zeroes, for each dispersion relation. It is worth noting that they are incompatible with the LO ChPT result, which is customarily used in the literature to implement simple fits to data.
\begin{table}[t] 
\caption{ Adler zero positions $\sqrt{s_A}$ (GeV), for the $I=1/2$ and $I=3/2$ $S$-waves  from dispersion relations.}
\vspace{0.3cm}
\centering 
\begin{tabular}{c c c c c} 
\hline
\rule[-0.2cm]{-0.1cm}{.55cm}  $I$ & \hspace{0.2cm} LO ChPT & \hspace{0.2cm} $\sqrt{s_{A_{FTPWDR}}}$ & $\sqrt{s_{A_{HPWDR}}}$ & $\sqrt{s_{A_{HPWDR_{sub}}}}$\\
\hline\hline  
\rule[-0.2cm]{-0.1cm}{.55cm} $1/2$ & $0.486$ & $0.466^{+0.006}_{-0.005}$ & $0.466^{+0.007}_{-0.005}$ & $0.470^{+0.010}_{-0.005}$\\
\rule[-0.2cm]{-0.1cm}{.55cm} $3/2$ & $0.516$ & $0.549^{+0.008}_{-0.0010}$ & $0.551^{+0.009}_{-0.0010}$ & $0.552^{+0.008}_{-0.010}$ \\
\hline
\end{tabular} 
\label{tab:adlerzeros} 
\end{table}

Finally, the fact that the low energy parameters can be obtained using an integral formula ensures their stability. Whereas for fits to the data the higher orders of the expansion must be obtained from derivatives, hence producing large spreads, the dispersion relations remain precise and accurate. This can be seen in Table~\ref{tab:leparameters}, where we provide our dispersive results for many different low energy parameters. It can be noted that even for higher angular momentums the parameters remain very precise and that the three dispersive solutions produce mostly identical values. When comparing our results to the previous dispersive calculation by the Paris group~\cite{Buettiker:2003pp} we observe compatible values most of the time. The most striking deviation appears for the $I=1/2$ $P$-wave values. Nonetheless, this tension is not surprising, as their dispersive prediction in that partial wave deviates from data. The comparison with the ChPT values from~\cite{Bijnens:2004bu, Bijnens:2014lea} shows larger tensions, however.

\begin{center}
\begin{table}[!h]
\centering
\footnotesize
\begin{tabular}{l|c|c|c|c|c|c}
\hline
&\multicolumn{4}{c|} {This work sum rules with CFD input}  &  Sum rules \cite{Buettiker:2003pp}&  NNLO ChPT  \\
 & FTPWDR & HPWDR & HPWDR$_{sub}$ & \textbf{Final Value} &  Fixed-$t$ &  \cite{Bijnens:2004bu} and 
\cite{Bijnens:2014lea}$^*$\\
\hline
\hline
\rule[-0.175cm]{-0.1cm}{.5cm} $m_\pi^3 b_0^{1/2}$ $\times$ 10 & 1.05$\pm$ 0.04 & 1.05$\pm$0.07 & 1.15$\pm$ 0.04 & \textbf{1.09$\pm$0.07}& 0.85$\pm$0.04 & 1.278    \\
\hline
\rule[-0.175cm]{-0.1cm}{.5cm} $m_\pi^3 b_0^{3/2}$ $\times$ 10& $-$0.43$\pm$0.02 & $-$0.41$\pm$0.03  &$-$0.45$\pm$0.02 & \textbf{$-$0.43$\pm$0.03}  & $-$0.37$\pm$0.03 & $-$0.326    \\
\hline
\rule[-0.175cm]{-0.1cm}{.5cm} $m_\pi^3 a_1^{1/2}$ $\times$ 10 & 0.228$\pm$0.010  & 0.218$\pm$0.008  & 0.222$\pm$0.006 & \textbf{0.222$\pm$0.009}  & 0.19$\pm$0.01  & 0.152       \\
\hline
\rule[-0.175cm]{-0.1cm}{.5cm} $m_\pi^5 b_1^{1/2}$ $\times$ 10$^{2}$ & 0.58$\pm$0.03 & 0.59$\pm$0.03  & 0.60$\pm$0.03 & \textbf{0.59$\pm$0.02}& 0.18$\pm$0.02   & 0.032      \\
\hline
\rule[-0.175cm]{-0.1cm}{.5cm} $m_\pi^3 a_1^{3/2}$ $\times$ 10$^{2}$ & 0.15$\pm$0.05 & 0.19$\pm$0.05  & 0.17$\pm$0.04 & \textbf{0.17$\pm$0.05} & 0.065$\pm$0.044   & 0.293         \\
\hline
\rule[-0.175cm]{-0.1cm}{.5cm} $m_\pi^5 b_1^{3/2}$ $\times$ 10$^{3}$ & $-$0.94$\pm$0.09 & $-$0.97$\pm$0.08 & $-$1.03$\pm$0.07 &  \textbf{$-$0.99$\pm$0.09}& $-$0.92$\pm$0.17  & 0.544         \\
\hline
\rule[-0.175cm]{-0.1cm}{.5cm} $m_\pi^5 a_2^{1/2}$ $\times$ 10$^{3}$ & 0.60$\pm$0.13  & 0.54$\pm$0.03   & 0.55$\pm$0.02  & \textbf{0.55$\pm$0.05} & 0.47$\pm$0.03   & 0.142        \\
\hline
\rule[-0.175cm]{-0.1cm}{.5cm} $m_\pi^7 b_2^{1/2}$ $\times$ 10$^{4}$ & $-$0.89$\pm$0.10  & $-$0.96$\pm$0.09   & $-$0.95$\pm$0.09  & \textbf{$-$0.94$\pm$0.09}  & $-$1.4$\pm$0.3   & $-$1.98       \\
\hline
\rule[-0.175cm]{-0.1cm}{.5cm} $m_\pi^5 a_2^{3/2}$ $\times$ 10$^{4}$ & $-$0.05$\pm$0.60  & $-$0.12$\pm$0.16  & $-$0.19$\pm$0.15 & \textbf{$-$0.15$\pm$0.18}  & $-$0.11$\pm$0.27   & $-$0.45         \\
\hline
\rule[-0.175cm]{-0.1cm}{.5cm} $m_\pi^7 b_2^{3/2}$ $\times$ 10$^{4}$ & $-$1.12$\pm$0.10  & $-$1.14$\pm$0.09   & $-$1.14$\pm$0.09 & \textbf{$-$1.13$\pm$0.06} & $-$0.96$\pm$0.26   & 0.61   \\
\hline
\end{tabular}
\caption{Determination of the $\pi K$ threshold parameters using our CFD  as input for the sum rules calculated from the partial-wave dispersion relations.  We also provide for comparison previous results using sum rules from Roy-Steiner equations by the Paris Group \cite{Buettiker:2003pp}. The last column lists NNLO ChPT results 
from \cite{Bijnens:2004bu,Bijnens:2014lea}. Our ``Final Value'' combines our three sum-rule results.}
\label{tab:leparameters}
\end{table}
\end{center}

\section{Summary}

Our recent analyses of low energy $\pi K$ interactions are summarized in this work. We use a combination of once subtracted and non-subtracted dispersion relations both for fixed-$t$ and hyperbolic sets. A high degree of precision is then obtained when constraining the available data. As a result, a set of powerful and simple parameterizations that fulfill all $S$-matrix principles is produced. These dispersion relations offer an advantageous way of extracting low energy parameters. Whereas for simple fits low energy parameters are obtained by unstable derivatives, these appear in the dispersive formalism as stable integral equations. We can thus extract many different observables with high precision and robustness. We summarize in this work our final determinations, including all uncertainty sources, and compare them to other previous extractions.

\section*{Acknowledgements}
 This project has received funding from the Spanish Ministerio de Ciencia e Innovación grant PID2019-106080GB-C21 and the European Union’s Horizon 2020 research and innovation program under grant agreement No 824093 (STRONG2020). AR acknowledges the financial support of the U.S. Department of Energy contract DE-SC0018416 at William \& Mary, and contract DE-AC05-06OR23177, under which Jefferson Science Associates, LLC, manages and operates Jefferson Lab. 

\bibliographystyle{JHEP}
\bibliography{largebiblio.bib}

\end{document}

%% file: figures/kappa12ufd.tex
\begingroup
  \makeatletter
  \providecommand\color[2][]{%
    \GenericError{(gnuplot) \space\space\space\@spaces}{%
      Package color not loaded in conjunction with
      terminal option `colourtext'%
    }{See the gnuplot documentation for explanation.%
    }{Either use 'blacktext' in gnuplot or load the package
      color.sty in LaTeX.}%
    \renewcommand\color[2][]{}%
  }%
  \providecommand\includegraphics[2][]{%
    \GenericError{(gnuplot) \space\space\space\@spaces}{%
      Package graphicx or graphics not loaded%
    }{See the gnuplot documentation for explanation.%
    }{The gnuplot epslatex terminal needs graphicx.sty or graphics.sty.}%
    \renewcommand\includegraphics[2][]{}%
  }%
  \providecommand\rotatebox[2]{#2}%
  \@ifundefined{ifGPcolor}{%
    \newif\ifGPcolor
    \GPcolortrue
  }{}%
  \@ifundefined{ifGPblacktext}{%
    \newif\ifGPblacktext
    \GPblacktexttrue
  }{}%
  \let\gplgaddtomacro\g@addto@macro
  \gdef\gplbacktext{}%
  \gdef\gplfronttext{}%
  \makeatother
  \ifGPblacktext
    \def\colorrgb#1{}%
    \def\colorgray#1{}%
  \else
    \ifGPcolor
      \def\colorrgb#1{\color[rgb]{#1}}%
      \def\colorgray#1{\color[gray]{#1}}%
      \expandafter\def\csname LTw\endcsname{\color{white}}%
      \expandafter\def\csname LTb\endcsname{\color{black}}%
      \expandafter\def\csname LTa\endcsname{\color{black}}%
      \expandafter\def\csname LT0\endcsname{\color[rgb]{1,0,0}}%
      \expandafter\def\csname LT1\endcsname{\color[rgb]{0,1,0}}%
      \expandafter\def\csname LT2\endcsname{\color[rgb]{0,0,1}}%
      \expandafter\def\csname LT3\endcsname{\color[rgb]{1,0,1}}%
      \expandafter\def\csname LT4\endcsname{\color[rgb]{0,1,1}}%
      \expandafter\def\csname LT5\endcsname{\color[rgb]{1,1,0}}%
      \expandafter\def\csname LT6\endcsname{\color[rgb]{0,0,0}}%
      \expandafter\def\csname LT7\endcsname{\color[rgb]{1,0.3,0}}%
      \expandafter\def\csname LT8\endcsname{\color[rgb]{0.5,0.5,0.5}}%
    \else
      \def\colorrgb#1{\color{black}}%
      \def\colorgray#1{\color[gray]{#1}}%
      \expandafter\def\csname LTw\endcsname{\color{white}}%
      \expandafter\def\csname LTb\endcsname{\color{black}}%
      \expandafter\def\csname LTa\endcsname{\color{black}}%
      \expandafter\def\csname LT0\endcsname{\color{black}}%
      \expandafter\def\csname LT1\endcsname{\color{black}}%
      \expandafter\def\csname LT2\endcsname{\color{black}}%
      \expandafter\def\csname LT3\endcsname{\color{black}}%
      \expandafter\def\csname LT4\endcsname{\color{black}}%
      \expandafter\def\csname LT5\endcsname{\color{black}}%
      \expandafter\def\csname LT6\endcsname{\color{black}}%
      \expandafter\def\csname LT7\endcsname{\color{black}}%
      \expandafter\def\csname LT8\endcsname{\color{black}}%
    \fi
  \fi
    \setlength{\unitlength}{0.0500bp}%
    \ifx\gptboxheight\undefined%
      \newlength{\gptboxheight}%
      \newlength{\gptboxwidth}%
      \newsavebox{\gptboxtext}%
    \fi%
    \setlength{\fboxrule}{0.5pt}%
    \setlength{\fboxsep}{1pt}%
\begin{picture}(7200.00,4032.00)%
    \gplgaddtomacro\gplbacktext{%
      \colorrgb{1.00,0.00,0.00}
      \put(228,806){\makebox(0,0)[r]{\strut{}$0.4$}}%
      \colorrgb{1.00,0.00,0.00}
      \put(228,1410){\makebox(0,0)[r]{\strut{}$0.5$}}%
      \colorrgb{1.00,0.00,0.00}
      \put(228,2015){\makebox(0,0)[r]{\strut{}$0.6$}}%
      \colorrgb{1.00,0.00,0.00}
      \put(228,2619){\makebox(0,0)[r]{\strut{}$0.7$}}%
      \colorrgb{1.00,0.00,0.00}
      \put(228,3223){\makebox(0,0)[r]{\strut{}$0.8$}}%
      \colorrgb{1.00,0.00,0.00}
      \put(228,3828){\makebox(0,0)[r]{\strut{}$0.9$}}%
      \colorrgb{1.00,0.00,0.00}
      \put(612,284){\makebox(0,0){\strut{}}}%
      \colorrgb{1.00,0.00,0.00}
      \put(1579,284){\makebox(0,0){\strut{}}}%
      \colorrgb{1.00,0.00,0.00}
      \put(2546,284){\makebox(0,0){\strut{}}}%
      \colorrgb{1.00,0.00,0.00}
      \put(3514,284){\makebox(0,0){\strut{}}}%
      \colorrgb{1.00,0.00,0.00}
      \put(4481,284){\makebox(0,0){\strut{}}}%
      \colorrgb{1.00,0.00,0.00}
      \put(5449,284){\makebox(0,0){\strut{}}}%
      \colorrgb{1.00,0.00,0.00}
      \put(6416,284){\makebox(0,0){\strut{}}}%
    }%
    \gplgaddtomacro\gplfronttext{%
      \csname LTb\endcsname
      \put(2868,1772){\makebox(0,0){\strut{}\Large{Re $f^{1/2}_0(s)$}}}%
      \colorrgb{1.00,0.00,0.00}
      \put(228,806){\makebox(0,0)[r]{\strut{}$0.4$}}%
      \colorrgb{1.00,0.00,0.00}
      \put(228,1410){\makebox(0,0)[r]{\strut{}$0.5$}}%
      \colorrgb{1.00,0.00,0.00}
      \put(228,2015){\makebox(0,0)[r]{\strut{}$0.6$}}%
      \colorrgb{1.00,0.00,0.00}
      \put(228,2619){\makebox(0,0)[r]{\strut{}$0.7$}}%
      \colorrgb{1.00,0.00,0.00}
      \put(228,3223){\makebox(0,0)[r]{\strut{}$0.8$}}%
      \colorrgb{1.00,0.00,0.00}
      \put(228,3828){\makebox(0,0)[r]{\strut{}$0.9$}}%
      \colorrgb{1.00,0.00,0.00}
      \put(612,284){\makebox(0,0){\strut{}}}%
      \colorrgb{1.00,0.00,0.00}
      \put(1579,284){\makebox(0,0){\strut{}}}%
      \colorrgb{1.00,0.00,0.00}
      \put(2546,284){\makebox(0,0){\strut{}}}%
      \colorrgb{1.00,0.00,0.00}
      \put(3514,284){\makebox(0,0){\strut{}}}%
      \colorrgb{1.00,0.00,0.00}
      \put(4481,284){\makebox(0,0){\strut{}}}%
      \colorrgb{1.00,0.00,0.00}
      \put(5449,284){\makebox(0,0){\strut{}}}%
      \colorrgb{1.00,0.00,0.00}
      \put(6416,284){\makebox(0,0){\strut{}}}%
      \csname LTb\endcsname
      \put(2256,3707){\makebox(0,0){\strut{}\Large{\textcolor{red}{$\hat{d}^2=5.2$}}}}%
      \put(1579,1108){\makebox(0,0){\strut{}\Large{\textcolor{blue}{$\hat{d}^2=13.7$}}}}%
      \put(6029,2166){\makebox(0,0){\strut{}\Large{\textcolor{orange}{$\hat{d}^2=1.5$}}}}%
      \put(998,3646){\makebox(0,0){\strut{}\huge{UFD}}}%
    }%
    \gplbacktext
    \put(0,0){\includegraphics{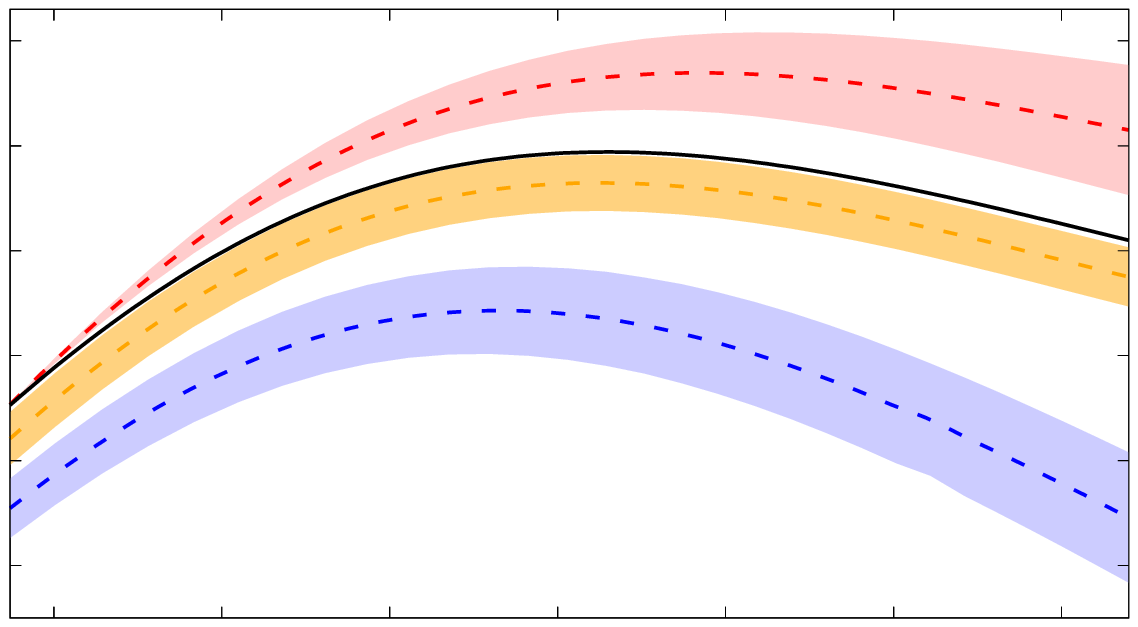}}%
    \gplfronttext
  \end{picture}%
\endgroup

%% file: figures/kappa12cfd.tex
\begingroup
  \makeatletter
  \providecommand\color[2][]{%
    \GenericError{(gnuplot) \space\space\space\@spaces}{%
      Package color not loaded in conjunction with
      terminal option `colourtext'%
    }{See the gnuplot documentation for explanation.%
    }{Either use 'blacktext' in gnuplot or load the package
      color.sty in LaTeX.}%
    \renewcommand\color[2][]{}%
  }%
  \providecommand\includegraphics[2][]{%
    \GenericError{(gnuplot) \space\space\space\@spaces}{%
      Package graphicx or graphics not loaded%
    }{See the gnuplot documentation for explanation.%
    }{The gnuplot epslatex terminal needs graphicx.sty or graphics.sty.}%
    \renewcommand\includegraphics[2][]{}%
  }%
  \providecommand\rotatebox[2]{#2}%
  \@ifundefined{ifGPcolor}{%
    \newif\ifGPcolor
    \GPcolortrue
  }{}%
  \@ifundefined{ifGPblacktext}{%
    \newif\ifGPblacktext
    \GPblacktexttrue
  }{}%
  \let\gplgaddtomacro\g@addto@macro
  \gdef\gplbacktext{}%
  \gdef\gplfronttext{}%
  \makeatother
  \ifGPblacktext
    \def\colorrgb#1{}%
    \def\colorgray#1{}%
  \else
    \ifGPcolor
      \def\colorrgb#1{\color[rgb]{#1}}%
      \def\colorgray#1{\color[gray]{#1}}%
      \expandafter\def\csname LTw\endcsname{\color{white}}%
      \expandafter\def\csname LTb\endcsname{\color{black}}%
      \expandafter\def\csname LTa\endcsname{\color{black}}%
      \expandafter\def\csname LT0\endcsname{\color[rgb]{1,0,0}}%
      \expandafter\def\csname LT1\endcsname{\color[rgb]{0,1,0}}%
      \expandafter\def\csname LT2\endcsname{\color[rgb]{0,0,1}}%
      \expandafter\def\csname LT3\endcsname{\color[rgb]{1,0,1}}%
      \expandafter\def\csname LT4\endcsname{\color[rgb]{0,1,1}}%
      \expandafter\def\csname LT5\endcsname{\color[rgb]{1,1,0}}%
      \expandafter\def\csname LT6\endcsname{\color[rgb]{0,0,0}}%
      \expandafter\def\csname LT7\endcsname{\color[rgb]{1,0.3,0}}%
      \expandafter\def\csname LT8\endcsname{\color[rgb]{0.5,0.5,0.5}}%
    \else
      \def\colorrgb#1{\color{black}}%
      \def\colorgray#1{\color[gray]{#1}}%
      \expandafter\def\csname LTw\endcsname{\color{white}}%
      \expandafter\def\csname LTb\endcsname{\color{black}}%
      \expandafter\def\csname LTa\endcsname{\color{black}}%
      \expandafter\def\csname LT0\endcsname{\color{black}}%
      \expandafter\def\csname LT1\endcsname{\color{black}}%
      \expandafter\def\csname LT2\endcsname{\color{black}}%
      \expandafter\def\csname LT3\endcsname{\color{black}}%
      \expandafter\def\csname LT4\endcsname{\color{black}}%
      \expandafter\def\csname LT5\endcsname{\color{black}}%
      \expandafter\def\csname LT6\endcsname{\color{black}}%
      \expandafter\def\csname LT7\endcsname{\color{black}}%
      \expandafter\def\csname LT8\endcsname{\color{black}}%
    \fi
  \fi
    \setlength{\unitlength}{0.0500bp}%
    \ifx\gptboxheight\undefined%
      \newlength{\gptboxheight}%
      \newlength{\gptboxwidth}%
      \newsavebox{\gptboxtext}%
    \fi%
    \setlength{\fboxrule}{0.5pt}%
    \setlength{\fboxsep}{1pt}%
\begin{picture}(7200.00,4032.00)%
    \gplgaddtomacro\gplbacktext{%
      \colorrgb{1.00,0.00,0.00}
      \put(228,806){\makebox(0,0)[r]{\strut{}}}%
      \colorrgb{1.00,0.00,0.00}
      \put(228,1410){\makebox(0,0)[r]{\strut{}}}%
      \colorrgb{1.00,0.00,0.00}
      \put(228,2015){\makebox(0,0)[r]{\strut{}}}%
      \colorrgb{1.00,0.00,0.00}
      \put(228,2619){\makebox(0,0)[r]{\strut{}}}%
      \colorrgb{1.00,0.00,0.00}
      \put(228,3223){\makebox(0,0)[r]{\strut{}}}%
      \colorrgb{1.00,0.00,0.00}
      \put(228,3828){\makebox(0,0)[r]{\strut{}}}%
      \colorrgb{1.00,0.00,0.00}
      \put(612,284){\makebox(0,0){\strut{}}}%
      \colorrgb{1.00,0.00,0.00}
      \put(1579,284){\makebox(0,0){\strut{}}}%
      \colorrgb{1.00,0.00,0.00}
      \put(2546,284){\makebox(0,0){\strut{}}}%
      \colorrgb{1.00,0.00,0.00}
      \put(3514,284){\makebox(0,0){\strut{}}}%
      \colorrgb{1.00,0.00,0.00}
      \put(4481,284){\makebox(0,0){\strut{}}}%
      \colorrgb{1.00,0.00,0.00}
      \put(5449,284){\makebox(0,0){\strut{}}}%
      \colorrgb{1.00,0.00,0.00}
      \put(6416,284){\makebox(0,0){\strut{}}}%
    }%
    \gplgaddtomacro\gplfronttext{%
      \csname LTb\endcsname
      \put(3264,1772){\makebox(0,0){\strut{}\Large{Re $f^{1/2}_0(s)$}}}%
      \csname LTb\endcsname
      \put(5816,1337){\makebox(0,0)[r]{\strut{}\large HPWDR$_{sub}$}}%
      \csname LTb\endcsname
      \put(5816,1117){\makebox(0,0)[r]{\strut{}\large HPWDR}}%
      \csname LTb\endcsname
      \put(5816,897){\makebox(0,0)[r]{\strut{}\large FTPWDR}}%
      \csname LTb\endcsname
      \put(5816,677){\makebox(0,0)[r]{\strut{}\large Input}}%
      \colorrgb{1.00,0.00,0.00}
      \put(228,806){\makebox(0,0)[r]{\strut{}}}%
      \colorrgb{1.00,0.00,0.00}
      \put(228,1410){\makebox(0,0)[r]{\strut{}}}%
      \colorrgb{1.00,0.00,0.00}
      \put(228,2015){\makebox(0,0)[r]{\strut{}}}%
      \colorrgb{1.00,0.00,0.00}
      \put(228,2619){\makebox(0,0)[r]{\strut{}}}%
      \colorrgb{1.00,0.00,0.00}
      \put(228,3223){\makebox(0,0)[r]{\strut{}}}%
      \colorrgb{1.00,0.00,0.00}
      \put(228,3828){\makebox(0,0)[r]{\strut{}}}%
      \colorrgb{1.00,0.00,0.00}
      \put(612,284){\makebox(0,0){\strut{}}}%
      \colorrgb{1.00,0.00,0.00}
      \put(1579,284){\makebox(0,0){\strut{}}}%
      \colorrgb{1.00,0.00,0.00}
      \put(2546,284){\makebox(0,0){\strut{}}}%
      \colorrgb{1.00,0.00,0.00}
      \put(3514,284){\makebox(0,0){\strut{}}}%
      \colorrgb{1.00,0.00,0.00}
      \put(4481,284){\makebox(0,0){\strut{}}}%
      \colorrgb{1.00,0.00,0.00}
      \put(5449,284){\makebox(0,0){\strut{}}}%
      \colorrgb{1.00,0.00,0.00}
      \put(6416,284){\makebox(0,0){\strut{}}}%
      \csname LTb\endcsname
      \put(2256,3707){\makebox(0,0){\strut{}\Large{\textcolor{red}{$\hat{d}^2=0.6$}}}}%
      \put(1579,1108){\makebox(0,0){\strut{}\Large{\textcolor{blue}{$\hat{d}^2=0.5$}}}}%
      \put(6029,3465){\makebox(0,0){\strut{}\Large{\textcolor{orange}{$\hat{d}^2=0.4$}}}}%
      \put(998,3646){\makebox(0,0){\strut{}\huge{CFD}}}%
    }%
    \gplbacktext
    \put(0,0){\includegraphics{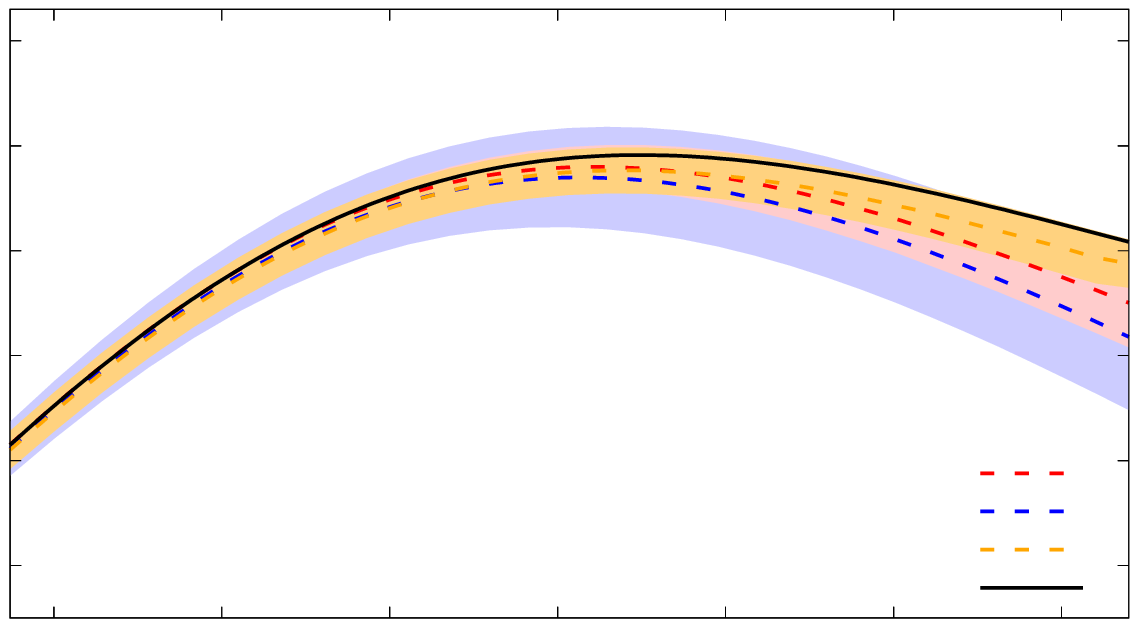}}%
    \gplfronttext
  \end{picture}%
\endgroup

%% file: figures/kappa32ufd.tex
\begingroup
  \makeatletter
  \providecommand\color[2][]{%
    \GenericError{(gnuplot) \space\space\space\@spaces}{%
      Package color not loaded in conjunction with
      terminal option `colourtext'%
    }{See the gnuplot documentation for explanation.%
    }{Either use 'blacktext' in gnuplot or load the package
      color.sty in LaTeX.}%
    \renewcommand\color[2][]{}%
  }%
  \providecommand\includegraphics[2][]{%
    \GenericError{(gnuplot) \space\space\space\@spaces}{%
      Package graphicx or graphics not loaded%
    }{See the gnuplot documentation for explanation.%
    }{The gnuplot epslatex terminal needs graphicx.sty or graphics.sty.}%
    \renewcommand\includegraphics[2][]{}%
  }%
  \providecommand\rotatebox[2]{#2}%
  \@ifundefined{ifGPcolor}{%
    \newif\ifGPcolor
    \GPcolortrue
  }{}%
  \@ifundefined{ifGPblacktext}{%
    \newif\ifGPblacktext
    \GPblacktexttrue
  }{}%
  \let\gplgaddtomacro\g@addto@macro
  \gdef\gplbacktext{}%
  \gdef\gplfronttext{}%
  \makeatother
  \ifGPblacktext
    \def\colorrgb#1{}%
    \def\colorgray#1{}%
  \else
    \ifGPcolor
      \def\colorrgb#1{\color[rgb]{#1}}%
      \def\colorgray#1{\color[gray]{#1}}%
      \expandafter\def\csname LTw\endcsname{\color{white}}%
      \expandafter\def\csname LTb\endcsname{\color{black}}%
      \expandafter\def\csname LTa\endcsname{\color{black}}%
      \expandafter\def\csname LT0\endcsname{\color[rgb]{1,0,0}}%
      \expandafter\def\csname LT1\endcsname{\color[rgb]{0,1,0}}%
      \expandafter\def\csname LT2\endcsname{\color[rgb]{0,0,1}}%
      \expandafter\def\csname LT3\endcsname{\color[rgb]{1,0,1}}%
      \expandafter\def\csname LT4\endcsname{\color[rgb]{0,1,1}}%
      \expandafter\def\csname LT5\endcsname{\color[rgb]{1,1,0}}%
      \expandafter\def\csname LT6\endcsname{\color[rgb]{0,0,0}}%
      \expandafter\def\csname LT7\endcsname{\color[rgb]{1,0.3,0}}%
      \expandafter\def\csname LT8\endcsname{\color[rgb]{0.5,0.5,0.5}}%
    \else
      \def\colorrgb#1{\color{black}}%
      \def\colorgray#1{\color[gray]{#1}}%
      \expandafter\def\csname LTw\endcsname{\color{white}}%
      \expandafter\def\csname LTb\endcsname{\color{black}}%
      \expandafter\def\csname LTa\endcsname{\color{black}}%
      \expandafter\def\csname LT0\endcsname{\color{black}}%
      \expandafter\def\csname LT1\endcsname{\color{black}}%
      \expandafter\def\csname LT2\endcsname{\color{black}}%
      \expandafter\def\csname LT3\endcsname{\color{black}}%
      \expandafter\def\csname LT4\endcsname{\color{black}}%
      \expandafter\def\csname LT5\endcsname{\color{black}}%
      \expandafter\def\csname LT6\endcsname{\color{black}}%
      \expandafter\def\csname LT7\endcsname{\color{black}}%
      \expandafter\def\csname LT8\endcsname{\color{black}}%
    \fi
  \fi
    \setlength{\unitlength}{0.0500bp}%
    \ifx\gptboxheight\undefined%
      \newlength{\gptboxheight}%
      \newlength{\gptboxwidth}%
      \newsavebox{\gptboxtext}%
    \fi%
    \setlength{\fboxrule}{0.5pt}%
    \setlength{\fboxsep}{1pt}%
\begin{picture}(7200.00,4032.00)%
    \gplgaddtomacro\gplbacktext{%
      \colorrgb{1.00,0.00,0.00}
      \put(228,504){\makebox(0,0)[r]{\strut{}$-0.5$}}%
      \colorrgb{1.00,0.00,0.00}
      \put(228,1205){\makebox(0,0)[r]{\strut{}$-0.4$}}%
      \colorrgb{1.00,0.00,0.00}
      \put(228,1906){\makebox(0,0)[r]{\strut{}$-0.3$}}%
      \colorrgb{1.00,0.00,0.00}
      \put(228,2607){\makebox(0,0)[r]{\strut{}$-0.2$}}%
      \colorrgb{1.00,0.00,0.00}
      \put(228,3308){\makebox(0,0)[r]{\strut{}$-0.1$}}%
      \colorrgb{1.00,0.00,0.00}
      \put(228,4009){\makebox(0,0)[r]{\strut{}$0$}}%
      \colorrgb{1.00,0.00,0.00}
      \put(612,284){\makebox(0,0){\strut{}}}%
      \colorrgb{1.00,0.00,0.00}
      \put(1579,284){\makebox(0,0){\strut{}}}%
      \colorrgb{1.00,0.00,0.00}
      \put(2546,284){\makebox(0,0){\strut{}}}%
      \colorrgb{1.00,0.00,0.00}
      \put(3514,284){\makebox(0,0){\strut{}}}%
      \colorrgb{1.00,0.00,0.00}
      \put(4481,284){\makebox(0,0){\strut{}}}%
      \colorrgb{1.00,0.00,0.00}
      \put(5449,284){\makebox(0,0){\strut{}}}%
      \colorrgb{1.00,0.00,0.00}
      \put(6416,284){\makebox(0,0){\strut{}}}%
    }%
    \gplgaddtomacro\gplfronttext{%
      \csname LTb\endcsname
      \put(4716,3576){\makebox(0,0){\strut{}\Large{Re $f^{3/2}_0(s)$}}}%
      \csname LTb\endcsname
      \put(2076,1337){\makebox(0,0)[r]{\strut{}\large HPWDR$_{sub}$}}%
      \csname LTb\endcsname
      \put(2076,1117){\makebox(0,0)[r]{\strut{}\large HPWDR}}%
      \csname LTb\endcsname
      \put(2076,897){\makebox(0,0)[r]{\strut{}\large FTPWDR}}%
      \csname LTb\endcsname
      \put(2076,677){\makebox(0,0)[r]{\strut{}\large Input}}%
      \colorrgb{1.00,0.00,0.00}
      \put(228,504){\makebox(0,0)[r]{\strut{}$-0.5$}}%
      \colorrgb{1.00,0.00,0.00}
      \put(228,1205){\makebox(0,0)[r]{\strut{}$-0.4$}}%
      \colorrgb{1.00,0.00,0.00}
      \put(228,1906){\makebox(0,0)[r]{\strut{}$-0.3$}}%
      \colorrgb{1.00,0.00,0.00}
      \put(228,2607){\makebox(0,0)[r]{\strut{}$-0.2$}}%
      \colorrgb{1.00,0.00,0.00}
      \put(228,3308){\makebox(0,0)[r]{\strut{}$-0.1$}}%
      \colorrgb{1.00,0.00,0.00}
      \put(228,4009){\makebox(0,0)[r]{\strut{}$0$}}%
      \colorrgb{1.00,0.00,0.00}
      \put(612,284){\makebox(0,0){\strut{}}}%
      \colorrgb{1.00,0.00,0.00}
      \put(1579,284){\makebox(0,0){\strut{}}}%
      \colorrgb{1.00,0.00,0.00}
      \put(2546,284){\makebox(0,0){\strut{}}}%
      \colorrgb{1.00,0.00,0.00}
      \put(3514,284){\makebox(0,0){\strut{}}}%
      \colorrgb{1.00,0.00,0.00}
      \put(4481,284){\makebox(0,0){\strut{}}}%
      \colorrgb{1.00,0.00,0.00}
      \put(5449,284){\makebox(0,0){\strut{}}}%
      \colorrgb{1.00,0.00,0.00}
      \put(6416,284){\makebox(0,0){\strut{}}}%
      \csname LTb\endcsname
      \put(3514,3448){\makebox(0,0){\strut{}\Large{\textcolor{red}{$\hat{d}^2=2.0$}}}}%
      \put(1579,3659){\makebox(0,0){\strut{}\Large{\textcolor{blue}{$\hat{d}^2=13.4$}}}}%
      \put(4481,2607){\makebox(0,0){\strut{}\Large{\textcolor{orange}{$\hat{d}^2=2.9$}}}}%
    }%
    \gplbacktext
    \put(0,0){\includegraphics{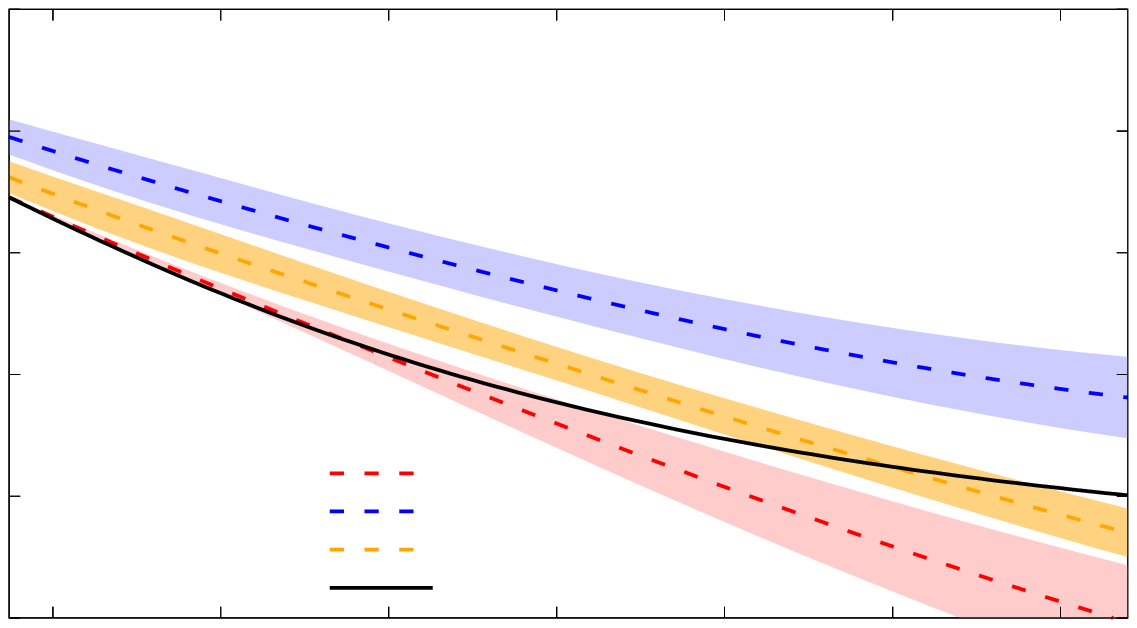}}%
    \gplfronttext
  \end{picture}%
\endgroup

%% file: figures/kappa32cfd.tex
\begingroup
  \makeatletter
  \providecommand\color[2][]{%
    \GenericError{(gnuplot) \space\space\space\@spaces}{%
      Package color not loaded in conjunction with
      terminal option `colourtext'%
    }{See the gnuplot documentation for explanation.%
    }{Either use 'blacktext' in gnuplot or load the package
      color.sty in LaTeX.}%
    \renewcommand\color[2][]{}%
  }%
  \providecommand\includegraphics[2][]{%
    \GenericError{(gnuplot) \space\space\space\@spaces}{%
      Package graphicx or graphics not loaded%
    }{See the gnuplot documentation for explanation.%
    }{The gnuplot epslatex terminal needs graphicx.sty or graphics.sty.}%
    \renewcommand\includegraphics[2][]{}%
  }%
  \providecommand\rotatebox[2]{#2}%
  \@ifundefined{ifGPcolor}{%
    \newif\ifGPcolor
    \GPcolortrue
  }{}%
  \@ifundefined{ifGPblacktext}{%
    \newif\ifGPblacktext
    \GPblacktexttrue
  }{}%
  \let\gplgaddtomacro\g@addto@macro
  \gdef\gplbacktext{}%
  \gdef\gplfronttext{}%
  \makeatother
  \ifGPblacktext
    \def\colorrgb#1{}%
    \def\colorgray#1{}%
  \else
    \ifGPcolor
      \def\colorrgb#1{\color[rgb]{#1}}%
      \def\colorgray#1{\color[gray]{#1}}%
      \expandafter\def\csname LTw\endcsname{\color{white}}%
      \expandafter\def\csname LTb\endcsname{\color{black}}%
      \expandafter\def\csname LTa\endcsname{\color{black}}%
      \expandafter\def\csname LT0\endcsname{\color[rgb]{1,0,0}}%
      \expandafter\def\csname LT1\endcsname{\color[rgb]{0,1,0}}%
      \expandafter\def\csname LT2\endcsname{\color[rgb]{0,0,1}}%
      \expandafter\def\csname LT3\endcsname{\color[rgb]{1,0,1}}%
      \expandafter\def\csname LT4\endcsname{\color[rgb]{0,1,1}}%
      \expandafter\def\csname LT5\endcsname{\color[rgb]{1,1,0}}%
      \expandafter\def\csname LT6\endcsname{\color[rgb]{0,0,0}}%
      \expandafter\def\csname LT7\endcsname{\color[rgb]{1,0.3,0}}%
      \expandafter\def\csname LT8\endcsname{\color[rgb]{0.5,0.5,0.5}}%
    \else
      \def\colorrgb#1{\color{black}}%
      \def\colorgray#1{\color[gray]{#1}}%
      \expandafter\def\csname LTw\endcsname{\color{white}}%
      \expandafter\def\csname LTb\endcsname{\color{black}}%
      \expandafter\def\csname LTa\endcsname{\color{black}}%
      \expandafter\def\csname LT0\endcsname{\color{black}}%
      \expandafter\def\csname LT1\endcsname{\color{black}}%
      \expandafter\def\csname LT2\endcsname{\color{black}}%
      \expandafter\def\csname LT3\endcsname{\color{black}}%
      \expandafter\def\csname LT4\endcsname{\color{black}}%
      \expandafter\def\csname LT5\endcsname{\color{black}}%
      \expandafter\def\csname LT6\endcsname{\color{black}}%
      \expandafter\def\csname LT7\endcsname{\color{black}}%
      \expandafter\def\csname LT8\endcsname{\color{black}}%
    \fi
  \fi
    \setlength{\unitlength}{0.0500bp}%
    \ifx\gptboxheight\undefined%
      \newlength{\gptboxheight}%
      \newlength{\gptboxwidth}%
      \newsavebox{\gptboxtext}%
    \fi%
    \setlength{\fboxrule}{0.5pt}%
    \setlength{\fboxsep}{1pt}%
\begin{picture}(7200.00,4032.00)%
    \gplgaddtomacro\gplbacktext{%
      \colorrgb{1.00,0.00,0.00}
      \put(228,504){\makebox(0,0)[r]{\strut{}}}%
      \colorrgb{1.00,0.00,0.00}
      \put(228,1205){\makebox(0,0)[r]{\strut{}}}%
      \colorrgb{1.00,0.00,0.00}
      \put(228,1906){\makebox(0,0)[r]{\strut{}}}%
      \colorrgb{1.00,0.00,0.00}
      \put(228,2607){\makebox(0,0)[r]{\strut{}}}%
      \colorrgb{1.00,0.00,0.00}
      \put(228,3308){\makebox(0,0)[r]{\strut{}}}%
      \colorrgb{1.00,0.00,0.00}
      \put(228,4009){\makebox(0,0)[r]{\strut{}}}%
      \colorrgb{1.00,0.00,0.00}
      \put(612,284){\makebox(0,0){\strut{}}}%
      \colorrgb{1.00,0.00,0.00}
      \put(1579,284){\makebox(0,0){\strut{}}}%
      \colorrgb{1.00,0.00,0.00}
      \put(2546,284){\makebox(0,0){\strut{}}}%
      \colorrgb{1.00,0.00,0.00}
      \put(3514,284){\makebox(0,0){\strut{}}}%
      \colorrgb{1.00,0.00,0.00}
      \put(4481,284){\makebox(0,0){\strut{}}}%
      \colorrgb{1.00,0.00,0.00}
      \put(5449,284){\makebox(0,0){\strut{}}}%
      \colorrgb{1.00,0.00,0.00}
      \put(6416,284){\makebox(0,0){\strut{}}}%
    }%
    \gplgaddtomacro\gplfronttext{%
      \csname LTb\endcsname
      \put(5244,3576){\makebox(0,0){\strut{}\Large{Re $f^{3/2}_0(s)$}}}%
      \csname LTb\endcsname
      \put(2076,1337){\makebox(0,0)[r]{\strut{}\large HPWDR$_{sub}$}}%
      \csname LTb\endcsname
      \put(2076,1117){\makebox(0,0)[r]{\strut{}\large HPWDR }}%
      \csname LTb\endcsname
      \put(2076,897){\makebox(0,0)[r]{\strut{}\large FTPWDR}}%
      \csname LTb\endcsname
      \put(2076,677){\makebox(0,0)[r]{\strut{}\large Input}}%
      \colorrgb{1.00,0.00,0.00}
      \put(228,504){\makebox(0,0)[r]{\strut{}}}%
      \colorrgb{1.00,0.00,0.00}
      \put(228,1205){\makebox(0,0)[r]{\strut{}}}%
      \colorrgb{1.00,0.00,0.00}
      \put(228,1906){\makebox(0,0)[r]{\strut{}}}%
      \colorrgb{1.00,0.00,0.00}
      \put(228,2607){\makebox(0,0)[r]{\strut{}}}%
      \colorrgb{1.00,0.00,0.00}
      \put(228,3308){\makebox(0,0)[r]{\strut{}}}%
      \colorrgb{1.00,0.00,0.00}
      \put(228,4009){\makebox(0,0)[r]{\strut{}}}%
      \colorrgb{1.00,0.00,0.00}
      \put(612,284){\makebox(0,0){\strut{}}}%
      \colorrgb{1.00,0.00,0.00}
      \put(1579,284){\makebox(0,0){\strut{}}}%
      \colorrgb{1.00,0.00,0.00}
      \put(2546,284){\makebox(0,0){\strut{}}}%
      \colorrgb{1.00,0.00,0.00}
      \put(3514,284){\makebox(0,0){\strut{}}}%
      \colorrgb{1.00,0.00,0.00}
      \put(4481,284){\makebox(0,0){\strut{}}}%
      \colorrgb{1.00,0.00,0.00}
      \put(5449,284){\makebox(0,0){\strut{}}}%
      \colorrgb{1.00,0.00,0.00}
      \put(6416,284){\makebox(0,0){\strut{}}}%
      \csname LTb\endcsname
      \put(3514,3448){\makebox(0,0){\strut{}\Large{\textcolor{red}{$\hat{d}^2=0.3$}}}}%
      \put(1579,3659){\makebox(0,0){\strut{}\Large{\textcolor{blue}{$\hat{d}^2=0.7$}}}}%
      \put(4481,2607){\makebox(0,0){\strut{}\Large{\textcolor{orange}{$\hat{d}^2=0.2$}}}}%
    }%
    \gplbacktext
    \put(0,0){\includegraphics{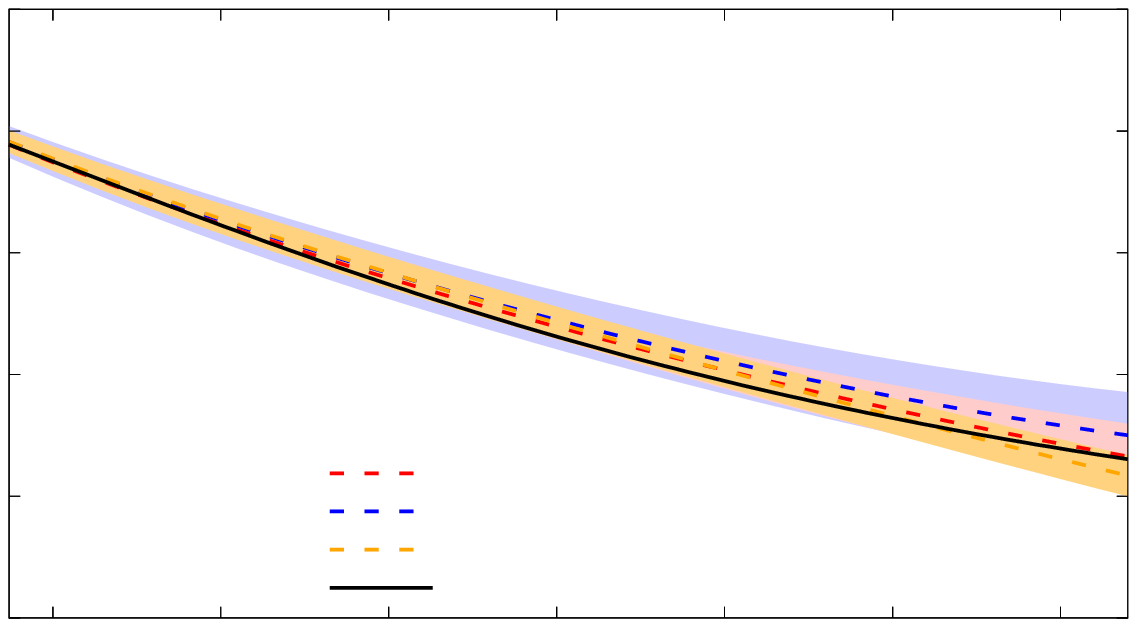}}%
    \gplfronttext
  \end{picture}%
\endgroup

%% file: figures/kappa121ufd.tex
\begingroup
  \makeatletter
  \providecommand\color[2][]{%
    \GenericError{(gnuplot) \space\space\space\@spaces}{%
      Package color not loaded in conjunction with
      terminal option `colourtext'%
    }{See the gnuplot documentation for explanation.%
    }{Either use 'blacktext' in gnuplot or load the package
      color.sty in LaTeX.}%
    \renewcommand\color[2][]{}%
  }%
  \providecommand\includegraphics[2][]{%
    \GenericError{(gnuplot) \space\space\space\@spaces}{%
      Package graphicx or graphics not loaded%
    }{See the gnuplot documentation for explanation.%
    }{The gnuplot epslatex terminal needs graphicx.sty or graphics.sty.}%
    \renewcommand\includegraphics[2][]{}%
  }%
  \providecommand\rotatebox[2]{#2}%
  \@ifundefined{ifGPcolor}{%
    \newif\ifGPcolor
    \GPcolortrue
  }{}%
  \@ifundefined{ifGPblacktext}{%
    \newif\ifGPblacktext
    \GPblacktexttrue
  }{}%
  \let\gplgaddtomacro\g@addto@macro
  \gdef\gplbacktext{}%
  \gdef\gplfronttext{}%
  \makeatother
  \ifGPblacktext
    \def\colorrgb#1{}%
    \def\colorgray#1{}%
  \else
    \ifGPcolor
      \def\colorrgb#1{\color[rgb]{#1}}%
      \def\colorgray#1{\color[gray]{#1}}%
      \expandafter\def\csname LTw\endcsname{\color{white}}%
      \expandafter\def\csname LTb\endcsname{\color{black}}%
      \expandafter\def\csname LTa\endcsname{\color{black}}%
      \expandafter\def\csname LT0\endcsname{\color[rgb]{1,0,0}}%
      \expandafter\def\csname LT1\endcsname{\color[rgb]{0,1,0}}%
      \expandafter\def\csname LT2\endcsname{\color[rgb]{0,0,1}}%
      \expandafter\def\csname LT3\endcsname{\color[rgb]{1,0,1}}%
      \expandafter\def\csname LT4\endcsname{\color[rgb]{0,1,1}}%
      \expandafter\def\csname LT5\endcsname{\color[rgb]{1,1,0}}%
      \expandafter\def\csname LT6\endcsname{\color[rgb]{0,0,0}}%
      \expandafter\def\csname LT7\endcsname{\color[rgb]{1,0.3,0}}%
      \expandafter\def\csname LT8\endcsname{\color[rgb]{0.5,0.5,0.5}}%
    \else
      \def\colorrgb#1{\color{black}}%
      \def\colorgray#1{\color[gray]{#1}}%
      \expandafter\def\csname LTw\endcsname{\color{white}}%
      \expandafter\def\csname LTb\endcsname{\color{black}}%
      \expandafter\def\csname LTa\endcsname{\color{black}}%
      \expandafter\def\csname LT0\endcsname{\color{black}}%
      \expandafter\def\csname LT1\endcsname{\color{black}}%
      \expandafter\def\csname LT2\endcsname{\color{black}}%
      \expandafter\def\csname LT3\endcsname{\color{black}}%
      \expandafter\def\csname LT4\endcsname{\color{black}}%
      \expandafter\def\csname LT5\endcsname{\color{black}}%
      \expandafter\def\csname LT6\endcsname{\color{black}}%
      \expandafter\def\csname LT7\endcsname{\color{black}}%
      \expandafter\def\csname LT8\endcsname{\color{black}}%
    \fi
  \fi
    \setlength{\unitlength}{0.0500bp}%
    \ifx\gptboxheight\undefined%
      \newlength{\gptboxheight}%
      \newlength{\gptboxwidth}%
      \newsavebox{\gptboxtext}%
    \fi%
    \setlength{\fboxrule}{0.5pt}%
    \setlength{\fboxsep}{1pt}%
\begin{picture}(7200.00,4032.00)%
    \gplgaddtomacro\gplbacktext{%
      \colorrgb{1.00,0.00,0.00}
      \put(228,504){\makebox(0,0)[r]{\strut{}$-1$}}%
      \colorrgb{1.00,0.00,0.00}
      \put(228,1380){\makebox(0,0)[r]{\strut{}$-0.5$}}%
      \colorrgb{1.00,0.00,0.00}
      \put(228,2257){\makebox(0,0)[r]{\strut{}$0$}}%
      \colorrgb{1.00,0.00,0.00}
      \put(228,3133){\makebox(0,0)[r]{\strut{}$0.5$}}%
      \colorrgb{1.00,0.00,0.00}
      \put(228,4009){\makebox(0,0)[r]{\strut{}$1$}}%
      \colorrgb{1.00,0.00,0.00}
      \put(612,284){\makebox(0,0){\strut{}}}%
      \colorrgb{1.00,0.00,0.00}
      \put(1579,284){\makebox(0,0){\strut{}}}%
      \colorrgb{1.00,0.00,0.00}
      \put(2546,284){\makebox(0,0){\strut{}}}%
      \colorrgb{1.00,0.00,0.00}
      \put(3514,284){\makebox(0,0){\strut{}}}%
      \colorrgb{1.00,0.00,0.00}
      \put(4481,284){\makebox(0,0){\strut{}}}%
      \colorrgb{1.00,0.00,0.00}
      \put(5449,284){\makebox(0,0){\strut{}}}%
      \colorrgb{1.00,0.00,0.00}
      \put(6416,284){\makebox(0,0){\strut{}}}%
    }%
    \gplgaddtomacro\gplfronttext{%
      \csname LTb\endcsname
      \put(4056,2036){\makebox(0,0){\strut{}\Large{Re $f^{1/2}_1(s)$}}}%
      \csname LTb\endcsname
      \put(2076,1337){\makebox(0,0)[r]{\strut{}\large HPWDR$_{sub}$}}%
      \csname LTb\endcsname
      \put(2076,1117){\makebox(0,0)[r]{\strut{}\large HPWDR }}%
      \csname LTb\endcsname
      \put(2076,897){\makebox(0,0)[r]{\strut{}\large FTPWDR}}%
      \csname LTb\endcsname
      \put(2076,677){\makebox(0,0)[r]{\strut{}\large Input}}%
      \colorrgb{1.00,0.00,0.00}
      \put(228,504){\makebox(0,0)[r]{\strut{}$-1$}}%
      \colorrgb{1.00,0.00,0.00}
      \put(228,1380){\makebox(0,0)[r]{\strut{}$-0.5$}}%
      \colorrgb{1.00,0.00,0.00}
      \put(228,2257){\makebox(0,0)[r]{\strut{}$0$}}%
      \colorrgb{1.00,0.00,0.00}
      \put(228,3133){\makebox(0,0)[r]{\strut{}$0.5$}}%
      \colorrgb{1.00,0.00,0.00}
      \put(228,4009){\makebox(0,0)[r]{\strut{}$1$}}%
      \colorrgb{1.00,0.00,0.00}
      \put(612,284){\makebox(0,0){\strut{}}}%
      \colorrgb{1.00,0.00,0.00}
      \put(1579,284){\makebox(0,0){\strut{}}}%
      \colorrgb{1.00,0.00,0.00}
      \put(2546,284){\makebox(0,0){\strut{}}}%
      \colorrgb{1.00,0.00,0.00}
      \put(3514,284){\makebox(0,0){\strut{}}}%
      \colorrgb{1.00,0.00,0.00}
      \put(4481,284){\makebox(0,0){\strut{}}}%
      \colorrgb{1.00,0.00,0.00}
      \put(5449,284){\makebox(0,0){\strut{}}}%
      \colorrgb{1.00,0.00,0.00}
      \put(6416,284){\makebox(0,0){\strut{}}}%
      \csname LTb\endcsname
      \put(3127,3746){\makebox(0,0){\strut{}\Large{\textcolor{red}{$\hat{d}^2=3.6$}}}}%
      \put(1385,3045){\makebox(0,0){\strut{}\Large{\textcolor{blue}{$\hat{d}^2=10.0$}}}}%
      \put(6029,3729){\makebox(0,0){\strut{}\Large{\textcolor{orange}{$\hat{d}^2=5.1$}}}}%
    }%
    \gplbacktext
    \put(0,0){\includegraphics{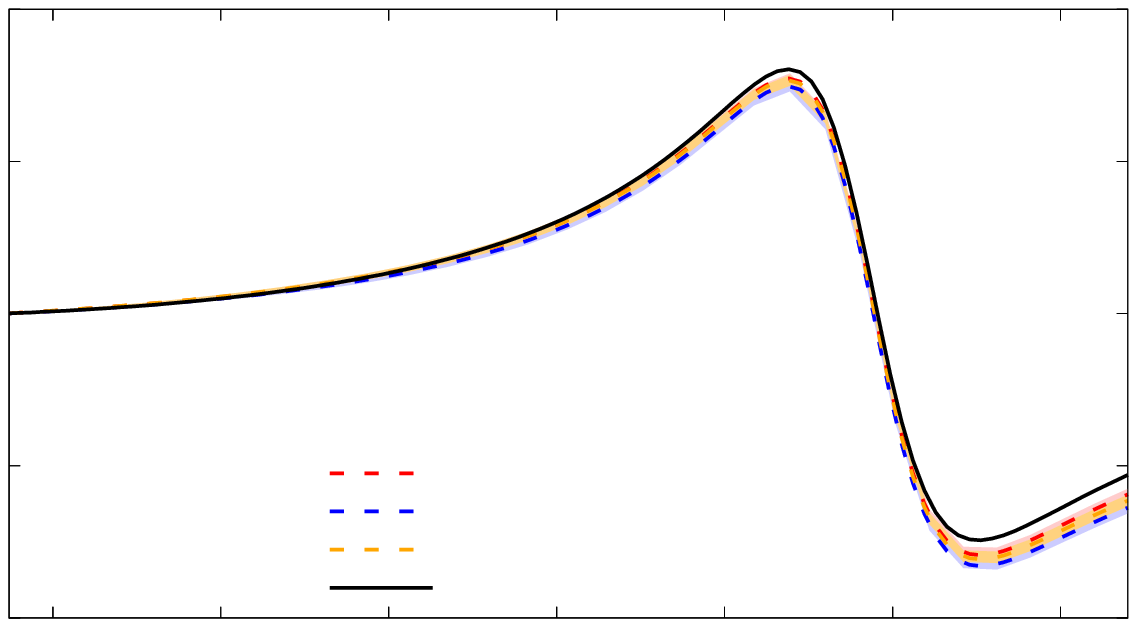}}%
    \gplfronttext
  \end{picture}%
\endgroup

%% file: figures/kappa121cfd.tex
\begingroup
  \makeatletter
  \providecommand\color[2][]{%
    \GenericError{(gnuplot) \space\space\space\@spaces}{%
      Package color not loaded in conjunction with
      terminal option `colourtext'%
    }{See the gnuplot documentation for explanation.%
    }{Either use 'blacktext' in gnuplot or load the package
      color.sty in LaTeX.}%
    \renewcommand\color[2][]{}%
  }%
  \providecommand\includegraphics[2][]{%
    \GenericError{(gnuplot) \space\space\space\@spaces}{%
      Package graphicx or graphics not loaded%
    }{See the gnuplot documentation for explanation.%
    }{The gnuplot epslatex terminal needs graphicx.sty or graphics.sty.}%
    \renewcommand\includegraphics[2][]{}%
  }%
  \providecommand\rotatebox[2]{#2}%
  \@ifundefined{ifGPcolor}{%
    \newif\ifGPcolor
    \GPcolortrue
  }{}%
  \@ifundefined{ifGPblacktext}{%
    \newif\ifGPblacktext
    \GPblacktexttrue
  }{}%
  \let\gplgaddtomacro\g@addto@macro
  \gdef\gplbacktext{}%
  \gdef\gplfronttext{}%
  \makeatother
  \ifGPblacktext
    \def\colorrgb#1{}%
    \def\colorgray#1{}%
  \else
    \ifGPcolor
      \def\colorrgb#1{\color[rgb]{#1}}%
      \def\colorgray#1{\color[gray]{#1}}%
      \expandafter\def\csname LTw\endcsname{\color{white}}%
      \expandafter\def\csname LTb\endcsname{\color{black}}%
      \expandafter\def\csname LTa\endcsname{\color{black}}%
      \expandafter\def\csname LT0\endcsname{\color[rgb]{1,0,0}}%
      \expandafter\def\csname LT1\endcsname{\color[rgb]{0,1,0}}%
      \expandafter\def\csname LT2\endcsname{\color[rgb]{0,0,1}}%
      \expandafter\def\csname LT3\endcsname{\color[rgb]{1,0,1}}%
      \expandafter\def\csname LT4\endcsname{\color[rgb]{0,1,1}}%
      \expandafter\def\csname LT5\endcsname{\color[rgb]{1,1,0}}%
      \expandafter\def\csname LT6\endcsname{\color[rgb]{0,0,0}}%
      \expandafter\def\csname LT7\endcsname{\color[rgb]{1,0.3,0}}%
      \expandafter\def\csname LT8\endcsname{\color[rgb]{0.5,0.5,0.5}}%
    \else
      \def\colorrgb#1{\color{black}}%
      \def\colorgray#1{\color[gray]{#1}}%
      \expandafter\def\csname LTw\endcsname{\color{white}}%
      \expandafter\def\csname LTb\endcsname{\color{black}}%
      \expandafter\def\csname LTa\endcsname{\color{black}}%
      \expandafter\def\csname LT0\endcsname{\color{black}}%
      \expandafter\def\csname LT1\endcsname{\color{black}}%
      \expandafter\def\csname LT2\endcsname{\color{black}}%
      \expandafter\def\csname LT3\endcsname{\color{black}}%
      \expandafter\def\csname LT4\endcsname{\color{black}}%
      \expandafter\def\csname LT5\endcsname{\color{black}}%
      \expandafter\def\csname LT6\endcsname{\color{black}}%
      \expandafter\def\csname LT7\endcsname{\color{black}}%
      \expandafter\def\csname LT8\endcsname{\color{black}}%
    \fi
  \fi
    \setlength{\unitlength}{0.0500bp}%
    \ifx\gptboxheight\undefined%
      \newlength{\gptboxheight}%
      \newlength{\gptboxwidth}%
      \newsavebox{\gptboxtext}%
    \fi%
    \setlength{\fboxrule}{0.5pt}%
    \setlength{\fboxsep}{1pt}%
\begin{picture}(7200.00,4032.00)%
    \gplgaddtomacro\gplbacktext{%
      \colorrgb{1.00,0.00,0.00}
      \put(228,504){\makebox(0,0)[r]{\strut{}}}%
      \colorrgb{1.00,0.00,0.00}
      \put(228,1380){\makebox(0,0)[r]{\strut{}}}%
      \colorrgb{1.00,0.00,0.00}
      \put(228,2257){\makebox(0,0)[r]{\strut{}}}%
      \colorrgb{1.00,0.00,0.00}
      \put(228,3133){\makebox(0,0)[r]{\strut{}}}%
      \colorrgb{1.00,0.00,0.00}
      \put(228,4009){\makebox(0,0)[r]{\strut{}}}%
      \colorrgb{1.00,0.00,0.00}
      \put(612,284){\makebox(0,0){\strut{}}}%
      \colorrgb{1.00,0.00,0.00}
      \put(1579,284){\makebox(0,0){\strut{}}}%
      \colorrgb{1.00,0.00,0.00}
      \put(2546,284){\makebox(0,0){\strut{}}}%
      \colorrgb{1.00,0.00,0.00}
      \put(3514,284){\makebox(0,0){\strut{}}}%
      \colorrgb{1.00,0.00,0.00}
      \put(4481,284){\makebox(0,0){\strut{}}}%
      \colorrgb{1.00,0.00,0.00}
      \put(5449,284){\makebox(0,0){\strut{}}}%
      \colorrgb{1.00,0.00,0.00}
      \put(6416,284){\makebox(0,0){\strut{}}}%
    }%
    \gplgaddtomacro\gplfronttext{%
      \csname LTb\endcsname
      \put(4584,2036){\makebox(0,0){\strut{}\Large{Re $f^{1/2}_1(s)$}}}%
      \csname LTb\endcsname
      \put(2076,1337){\makebox(0,0)[r]{\strut{}\large HPWDR$_{sub}$}}%
      \csname LTb\endcsname
      \put(2076,1117){\makebox(0,0)[r]{\strut{}\large HPWDR }}%
      \csname LTb\endcsname
      \put(2076,897){\makebox(0,0)[r]{\strut{}\large FTPWDR}}%
      \csname LTb\endcsname
      \put(2076,677){\makebox(0,0)[r]{\strut{}\large Input}}%
      \colorrgb{1.00,0.00,0.00}
      \put(228,504){\makebox(0,0)[r]{\strut{}}}%
      \colorrgb{1.00,0.00,0.00}
      \put(228,1380){\makebox(0,0)[r]{\strut{}}}%
      \colorrgb{1.00,0.00,0.00}
      \put(228,2257){\makebox(0,0)[r]{\strut{}}}%
      \colorrgb{1.00,0.00,0.00}
      \put(228,3133){\makebox(0,0)[r]{\strut{}}}%
      \colorrgb{1.00,0.00,0.00}
      \put(228,4009){\makebox(0,0)[r]{\strut{}}}%
      \colorrgb{1.00,0.00,0.00}
      \put(612,284){\makebox(0,0){\strut{}}}%
      \colorrgb{1.00,0.00,0.00}
      \put(1579,284){\makebox(0,0){\strut{}}}%
      \colorrgb{1.00,0.00,0.00}
      \put(2546,284){\makebox(0,0){\strut{}}}%
      \colorrgb{1.00,0.00,0.00}
      \put(3514,284){\makebox(0,0){\strut{}}}%
      \colorrgb{1.00,0.00,0.00}
      \put(4481,284){\makebox(0,0){\strut{}}}%
      \colorrgb{1.00,0.00,0.00}
      \put(5449,284){\makebox(0,0){\strut{}}}%
      \colorrgb{1.00,0.00,0.00}
      \put(6416,284){\makebox(0,0){\strut{}}}%
      \csname LTb\endcsname
      \put(3127,3746){\makebox(0,0){\strut{}\Large{\textcolor{red}{$\hat{d}^2=0.2$}}}}%
      \put(1385,3045){\makebox(0,0){\strut{}\Large{\textcolor{blue}{$\hat{d}^2=0.5$}}}}%
      \put(6029,3729){\makebox(0,0){\strut{}\Large{\textcolor{orange}{$\hat{d}^2=0.5$}}}}%
    }%
    \gplbacktext
    \put(0,0){\includegraphics{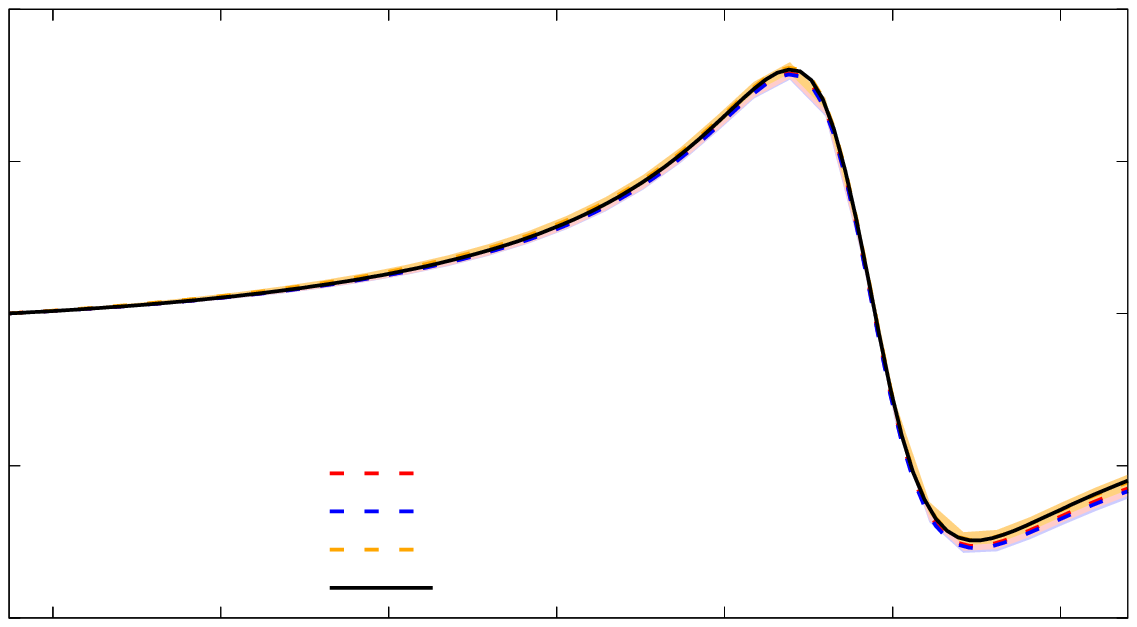}}%
    \gplfronttext
  \end{picture}%
\endgroup

%% file: figures/kappa321ufd.tex
\begingroup
  \makeatletter
  \providecommand\color[2][]{%
    \GenericError{(gnuplot) \space\space\space\@spaces}{%
      Package color not loaded in conjunction with
      terminal option `colourtext'%
    }{See the gnuplot documentation for explanation.%
    }{Either use 'blacktext' in gnuplot or load the package
      color.sty in LaTeX.}%
    \renewcommand\color[2][]{}%
  }%
  \providecommand\includegraphics[2][]{%
    \GenericError{(gnuplot) \space\space\space\@spaces}{%
      Package graphicx or graphics not loaded%
    }{See the gnuplot documentation for explanation.%
    }{The gnuplot epslatex terminal needs graphicx.sty or graphics.sty.}%
    \renewcommand\includegraphics[2][]{}%
  }%
  \providecommand\rotatebox[2]{#2}%
  \@ifundefined{ifGPcolor}{%
    \newif\ifGPcolor
    \GPcolortrue
  }{}%
  \@ifundefined{ifGPblacktext}{%
    \newif\ifGPblacktext
    \GPblacktexttrue
  }{}%
  \let\gplgaddtomacro\g@addto@macro
  \gdef\gplbacktext{}%
  \gdef\gplfronttext{}%
  \makeatother
  \ifGPblacktext
    \def\colorrgb#1{}%
    \def\colorgray#1{}%
  \else
    \ifGPcolor
      \def\colorrgb#1{\color[rgb]{#1}}%
      \def\colorgray#1{\color[gray]{#1}}%
      \expandafter\def\csname LTw\endcsname{\color{white}}%
      \expandafter\def\csname LTb\endcsname{\color{black}}%
      \expandafter\def\csname LTa\endcsname{\color{black}}%
      \expandafter\def\csname LT0\endcsname{\color[rgb]{1,0,0}}%
      \expandafter\def\csname LT1\endcsname{\color[rgb]{0,1,0}}%
      \expandafter\def\csname LT2\endcsname{\color[rgb]{0,0,1}}%
      \expandafter\def\csname LT3\endcsname{\color[rgb]{1,0,1}}%
      \expandafter\def\csname LT4\endcsname{\color[rgb]{0,1,1}}%
      \expandafter\def\csname LT5\endcsname{\color[rgb]{1,1,0}}%
      \expandafter\def\csname LT6\endcsname{\color[rgb]{0,0,0}}%
      \expandafter\def\csname LT7\endcsname{\color[rgb]{1,0.3,0}}%
      \expandafter\def\csname LT8\endcsname{\color[rgb]{0.5,0.5,0.5}}%
    \else
      \def\colorrgb#1{\color{black}}%
      \def\colorgray#1{\color[gray]{#1}}%
      \expandafter\def\csname LTw\endcsname{\color{white}}%
      \expandafter\def\csname LTb\endcsname{\color{black}}%
      \expandafter\def\csname LTa\endcsname{\color{black}}%
      \expandafter\def\csname LT0\endcsname{\color{black}}%
      \expandafter\def\csname LT1\endcsname{\color{black}}%
      \expandafter\def\csname LT2\endcsname{\color{black}}%
      \expandafter\def\csname LT3\endcsname{\color{black}}%
      \expandafter\def\csname LT4\endcsname{\color{black}}%
      \expandafter\def\csname LT5\endcsname{\color{black}}%
      \expandafter\def\csname LT6\endcsname{\color{black}}%
      \expandafter\def\csname LT7\endcsname{\color{black}}%
      \expandafter\def\csname LT8\endcsname{\color{black}}%
    \fi
  \fi
    \setlength{\unitlength}{0.0500bp}%
    \ifx\gptboxheight\undefined%
      \newlength{\gptboxheight}%
      \newlength{\gptboxwidth}%
      \newsavebox{\gptboxtext}%
    \fi%
    \setlength{\fboxrule}{0.5pt}%
    \setlength{\fboxsep}{1pt}%
\begin{picture}(7200.00,4032.00)%
    \gplgaddtomacro\gplbacktext{%
      \colorrgb{1.00,0.00,0.00}
      \put(228,504){\makebox(0,0)[r]{\strut{}$-0.1$}}%
      \colorrgb{1.00,0.00,0.00}
      \put(228,1380){\makebox(0,0)[r]{\strut{}$-0.05$}}%
      \colorrgb{1.00,0.00,0.00}
      \put(228,2257){\makebox(0,0)[r]{\strut{}$0$}}%
      \colorrgb{1.00,0.00,0.00}
      \put(228,3133){\makebox(0,0)[r]{\strut{}$0.05$}}%
      \colorrgb{1.00,0.00,0.00}
      \put(228,4009){\makebox(0,0)[r]{\strut{}$0.1$}}%
      \colorrgb{1.00,0.00,0.00}
      \put(612,284){\makebox(0,0){\strut{}$0.65$}}%
      \colorrgb{1.00,0.00,0.00}
      \put(1579,284){\makebox(0,0){\strut{}$0.7$}}%
      \colorrgb{1.00,0.00,0.00}
      \put(2546,284){\makebox(0,0){\strut{}$0.75$}}%
      \colorrgb{1.00,0.00,0.00}
      \put(3514,284){\makebox(0,0){\strut{}$0.8$}}%
      \colorrgb{1.00,0.00,0.00}
      \put(4481,284){\makebox(0,0){\strut{}$0.85$}}%
      \colorrgb{1.00,0.00,0.00}
      \put(5449,284){\makebox(0,0){\strut{}$0.9$}}%
      \colorrgb{1.00,0.00,0.00}
      \put(6416,284){\makebox(0,0){\strut{}$0.95$}}%
    }%
    \gplgaddtomacro\gplfronttext{%
      \csname LTb\endcsname
      \put(3924,3356){\makebox(0,0){\strut{}\Large{Re $f^{3/2}_1(s)$}}}%
      \put(3581,-46){\makebox(0,0){\strut{}\large{$\sqrt{s}$ GeV}}}%
      \csname LTb\endcsname
      \put(2076,1337){\makebox(0,0)[r]{\strut{}\large HPWDR$_{sub}$}}%
      \csname LTb\endcsname
      \put(2076,1117){\makebox(0,0)[r]{\strut{}\large HPWDR}}%
      \csname LTb\endcsname
      \put(2076,897){\makebox(0,0)[r]{\strut{}\large FTPWDR}}%
      \csname LTb\endcsname
      \put(2076,677){\makebox(0,0)[r]{\strut{}\large Input}}%
      \colorrgb{1.00,0.00,0.00}
      \put(228,504){\makebox(0,0)[r]{\strut{}$-0.1$}}%
      \colorrgb{1.00,0.00,0.00}
      \put(228,1380){\makebox(0,0)[r]{\strut{}$-0.05$}}%
      \colorrgb{1.00,0.00,0.00}
      \put(228,2257){\makebox(0,0)[r]{\strut{}$0$}}%
      \colorrgb{1.00,0.00,0.00}
      \put(228,3133){\makebox(0,0)[r]{\strut{}$0.05$}}%
      \colorrgb{1.00,0.00,0.00}
      \put(228,4009){\makebox(0,0)[r]{\strut{}$0.1$}}%
      \colorrgb{1.00,0.00,0.00}
      \put(612,284){\makebox(0,0){\strut{}$0.65$}}%
      \colorrgb{1.00,0.00,0.00}
      \put(1579,284){\makebox(0,0){\strut{}$0.7$}}%
      \colorrgb{1.00,0.00,0.00}
      \put(2546,284){\makebox(0,0){\strut{}$0.75$}}%
      \colorrgb{1.00,0.00,0.00}
      \put(3514,284){\makebox(0,0){\strut{}$0.8$}}%
      \colorrgb{1.00,0.00,0.00}
      \put(4481,284){\makebox(0,0){\strut{}$0.85$}}%
      \colorrgb{1.00,0.00,0.00}
      \put(5449,284){\makebox(0,0){\strut{}$0.9$}}%
      \colorrgb{1.00,0.00,0.00}
      \put(6416,284){\makebox(0,0){\strut{}$0.95$}}%
      \csname LTb\endcsname
      \put(1579,3133){\makebox(0,0){\strut{}\Large{\textcolor{red}{$\hat{d}^2=0.1$}}}}%
      \put(3514,1556){\makebox(0,0){\strut{}\Large{\textcolor{blue}{$\hat{d}^2=8.9$}}}}%
      \put(5449,1117){\makebox(0,0){\strut{}\Large{\textcolor{orange}{$\hat{d}^2=0.5$}}}}%
    }%
    \gplbacktext
    \put(0,0){\includegraphics{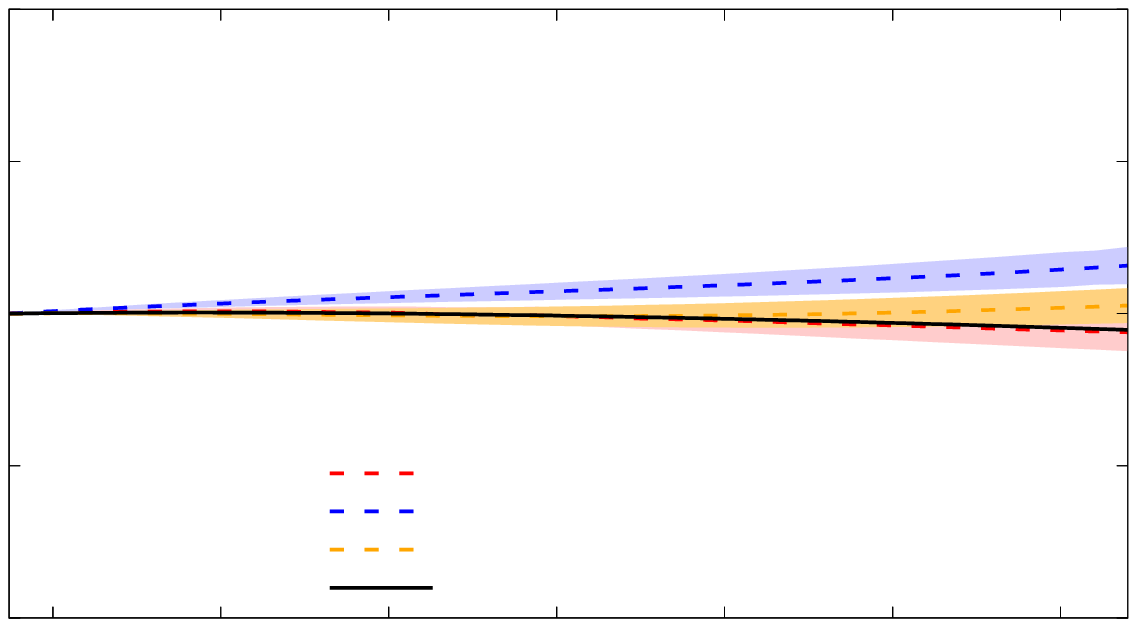}}%
    \gplfronttext
  \end{picture}%
\endgroup

%% file: figures/kappa321cfd.tex
\begingroup
  \makeatletter
  \providecommand\color[2][]{%
    \GenericError{(gnuplot) \space\space\space\@spaces}{%
      Package color not loaded in conjunction with
      terminal option `colourtext'%
    }{See the gnuplot documentation for explanation.%
    }{Either use 'blacktext' in gnuplot or load the package
      color.sty in LaTeX.}%
    \renewcommand\color[2][]{}%
  }%
  \providecommand\includegraphics[2][]{%
    \GenericError{(gnuplot) \space\space\space\@spaces}{%
      Package graphicx or graphics not loaded%
    }{See the gnuplot documentation for explanation.%
    }{The gnuplot epslatex terminal needs graphicx.sty or graphics.sty.}%
    \renewcommand\includegraphics[2][]{}%
  }%
  \providecommand\rotatebox[2]{#2}%
  \@ifundefined{ifGPcolor}{%
    \newif\ifGPcolor
    \GPcolortrue
  }{}%
  \@ifundefined{ifGPblacktext}{%
    \newif\ifGPblacktext
    \GPblacktexttrue
  }{}%
  \let\gplgaddtomacro\g@addto@macro
  \gdef\gplbacktext{}%
  \gdef\gplfronttext{}%
  \makeatother
  \ifGPblacktext
    \def\colorrgb#1{}%
    \def\colorgray#1{}%
  \else
    \ifGPcolor
      \def\colorrgb#1{\color[rgb]{#1}}%
      \def\colorgray#1{\color[gray]{#1}}%
      \expandafter\def\csname LTw\endcsname{\color{white}}%
      \expandafter\def\csname LTb\endcsname{\color{black}}%
      \expandafter\def\csname LTa\endcsname{\color{black}}%
      \expandafter\def\csname LT0\endcsname{\color[rgb]{1,0,0}}%
      \expandafter\def\csname LT1\endcsname{\color[rgb]{0,1,0}}%
      \expandafter\def\csname LT2\endcsname{\color[rgb]{0,0,1}}%
      \expandafter\def\csname LT3\endcsname{\color[rgb]{1,0,1}}%
      \expandafter\def\csname LT4\endcsname{\color[rgb]{0,1,1}}%
      \expandafter\def\csname LT5\endcsname{\color[rgb]{1,1,0}}%
      \expandafter\def\csname LT6\endcsname{\color[rgb]{0,0,0}}%
      \expandafter\def\csname LT7\endcsname{\color[rgb]{1,0.3,0}}%
      \expandafter\def\csname LT8\endcsname{\color[rgb]{0.5,0.5,0.5}}%
    \else
      \def\colorrgb#1{\color{black}}%
      \def\colorgray#1{\color[gray]{#1}}%
      \expandafter\def\csname LTw\endcsname{\color{white}}%
      \expandafter\def\csname LTb\endcsname{\color{black}}%
      \expandafter\def\csname LTa\endcsname{\color{black}}%
      \expandafter\def\csname LT0\endcsname{\color{black}}%
      \expandafter\def\csname LT1\endcsname{\color{black}}%
      \expandafter\def\csname LT2\endcsname{\color{black}}%
      \expandafter\def\csname LT3\endcsname{\color{black}}%
      \expandafter\def\csname LT4\endcsname{\color{black}}%
      \expandafter\def\csname LT5\endcsname{\color{black}}%
      \expandafter\def\csname LT6\endcsname{\color{black}}%
      \expandafter\def\csname LT7\endcsname{\color{black}}%
      \expandafter\def\csname LT8\endcsname{\color{black}}%
    \fi
  \fi
    \setlength{\unitlength}{0.0500bp}%
    \ifx\gptboxheight\undefined%
      \newlength{\gptboxheight}%
      \newlength{\gptboxwidth}%
      \newsavebox{\gptboxtext}%
    \fi%
    \setlength{\fboxrule}{0.5pt}%
    \setlength{\fboxsep}{1pt}%
\begin{picture}(7200.00,4032.00)%
    \gplgaddtomacro\gplbacktext{%
      \colorrgb{1.00,0.00,0.00}
      \put(228,504){\makebox(0,0)[r]{\strut{}}}%
      \colorrgb{1.00,0.00,0.00}
      \put(228,1380){\makebox(0,0)[r]{\strut{}}}%
      \colorrgb{1.00,0.00,0.00}
      \put(228,2257){\makebox(0,0)[r]{\strut{}}}%
      \colorrgb{1.00,0.00,0.00}
      \put(228,3133){\makebox(0,0)[r]{\strut{}}}%
      \colorrgb{1.00,0.00,0.00}
      \put(228,4009){\makebox(0,0)[r]{\strut{}}}%
      \colorrgb{1.00,0.00,0.00}
      \put(612,284){\makebox(0,0){\strut{}$0.65$}}%
      \colorrgb{1.00,0.00,0.00}
      \put(1579,284){\makebox(0,0){\strut{}$0.7$}}%
      \colorrgb{1.00,0.00,0.00}
      \put(2546,284){\makebox(0,0){\strut{}$0.75$}}%
      \colorrgb{1.00,0.00,0.00}
      \put(3514,284){\makebox(0,0){\strut{}$0.8$}}%
      \colorrgb{1.00,0.00,0.00}
      \put(4481,284){\makebox(0,0){\strut{}$0.85$}}%
      \colorrgb{1.00,0.00,0.00}
      \put(5449,284){\makebox(0,0){\strut{}$0.9$}}%
      \colorrgb{1.00,0.00,0.00}
      \put(6416,284){\makebox(0,0){\strut{}$0.95$}}%
    }%
    \gplgaddtomacro\gplfronttext{%
      \csname LTb\endcsname
      \put(4584,3356){\makebox(0,0){\strut{}\Large{Re $f^{3/2}_1(s)$}}}%
      \put(3581,-46){\makebox(0,0){\strut{}\large{$\sqrt{s}$ GeV}}}%
      \csname LTb\endcsname
      \put(2076,1337){\makebox(0,0)[r]{\strut{}\large HPWDR$_{sub}$}}%
      \csname LTb\endcsname
      \put(2076,1117){\makebox(0,0)[r]{\strut{}\large HDR }}%
      \csname LTb\endcsname
      \put(2076,897){\makebox(0,0)[r]{\strut{}\large FTPWDR}}%
      \csname LTb\endcsname
      \put(2076,677){\makebox(0,0)[r]{\strut{}\large Input}}%
      \colorrgb{1.00,0.00,0.00}
      \put(228,504){\makebox(0,0)[r]{\strut{}}}%
      \colorrgb{1.00,0.00,0.00}
      \put(228,1380){\makebox(0,0)[r]{\strut{}}}%
      \colorrgb{1.00,0.00,0.00}
      \put(228,2257){\makebox(0,0)[r]{\strut{}}}%
      \colorrgb{1.00,0.00,0.00}
      \put(228,3133){\makebox(0,0)[r]{\strut{}}}%
      \colorrgb{1.00,0.00,0.00}
      \put(228,4009){\makebox(0,0)[r]{\strut{}}}%
      \colorrgb{1.00,0.00,0.00}
      \put(612,284){\makebox(0,0){\strut{}$0.65$}}%
      \colorrgb{1.00,0.00,0.00}
      \put(1579,284){\makebox(0,0){\strut{}$0.7$}}%
      \colorrgb{1.00,0.00,0.00}
      \put(2546,284){\makebox(0,0){\strut{}$0.75$}}%
      \colorrgb{1.00,0.00,0.00}
      \put(3514,284){\makebox(0,0){\strut{}$0.8$}}%
      \colorrgb{1.00,0.00,0.00}
      \put(4481,284){\makebox(0,0){\strut{}$0.85$}}%
      \colorrgb{1.00,0.00,0.00}
      \put(5449,284){\makebox(0,0){\strut{}$0.9$}}%
      \colorrgb{1.00,0.00,0.00}
      \put(6416,284){\makebox(0,0){\strut{}$0.95$}}%
      \csname LTb\endcsname
      \put(1579,3133){\makebox(0,0){\strut{}\Large{\textcolor{red}{$\hat{d}^2=0.1$}}}}%
      \put(3514,1556){\makebox(0,0){\strut{}\Large{\textcolor{blue}{$\hat{d}^2=0.9$}}}}%
      \put(5449,1117){\makebox(0,0){\strut{}\Large{\textcolor{orange}{$\hat{d}^2=0.8$}}}}%
    }%
    \gplbacktext
    \put(0,0){\includegraphics{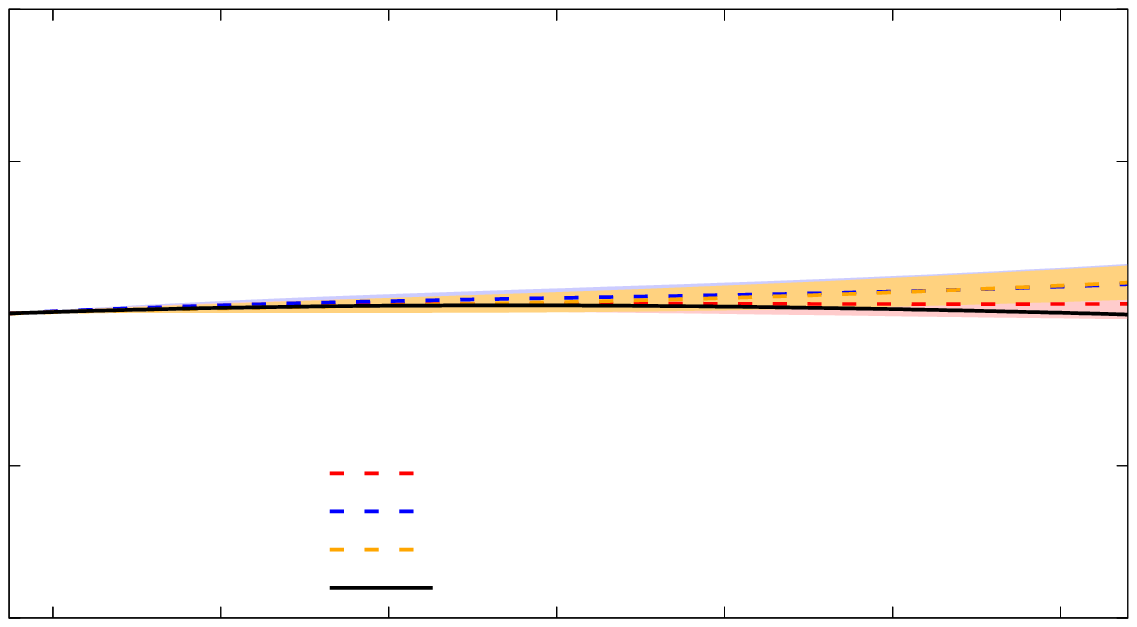}}%
    \gplfronttext
  \end{picture}%
\endgroup

%% file: figures/selas32cfd.tex
\begingroup
  \makeatletter
  \providecommand\color[2][]{%
    \GenericError{(gnuplot) \space\space\space\@spaces}{%
      Package color not loaded in conjunction with
      terminal option `colourtext'%
    }{See the gnuplot documentation for explanation.%
    }{Either use 'blacktext' in gnuplot or load the package
      color.sty in LaTeX.}%
    \renewcommand\color[2][]{}%
  }%
  \providecommand\includegraphics[2][]{%
    \GenericError{(gnuplot) \space\space\space\@spaces}{%
      Package graphicx or graphics not loaded%
    }{See the gnuplot documentation for explanation.%
    }{The gnuplot epslatex terminal needs graphicx.sty or graphics.sty.}%
    \renewcommand\includegraphics[2][]{}%
  }%
  \providecommand\rotatebox[2]{#2}%
  \@ifundefined{ifGPcolor}{%
    \newif\ifGPcolor
    \GPcolorfalse
  }{}%
  \@ifundefined{ifGPblacktext}{%
    \newif\ifGPblacktext
    \GPblacktexttrue
  }{}%
  \let\gplgaddtomacro\g@addto@macro
  \gdef\gplbacktext{}%
  \gdef\gplfronttext{}%
  \makeatother
  \ifGPblacktext
    \def\colorrgb#1{}%
    \def\colorgray#1{}%
  \else
    \ifGPcolor
      \def\colorrgb#1{\color[rgb]{#1}}%
      \def\colorgray#1{\color[gray]{#1}}%
      \expandafter\def\csname LTw\endcsname{\color{white}}%
      \expandafter\def\csname LTb\endcsname{\color{black}}%
      \expandafter\def\csname LTa\endcsname{\color{black}}%
      \expandafter\def\csname LT0\endcsname{\color[rgb]{1,0,0}}%
      \expandafter\def\csname LT1\endcsname{\color[rgb]{0,1,0}}%
      \expandafter\def\csname LT2\endcsname{\color[rgb]{0,0,1}}%
      \expandafter\def\csname LT3\endcsname{\color[rgb]{1,0,1}}%
      \expandafter\def\csname LT4\endcsname{\color[rgb]{0,1,1}}%
      \expandafter\def\csname LT5\endcsname{\color[rgb]{1,1,0}}%
      \expandafter\def\csname LT6\endcsname{\color[rgb]{0,0,0}}%
      \expandafter\def\csname LT7\endcsname{\color[rgb]{1,0.3,0}}%
      \expandafter\def\csname LT8\endcsname{\color[rgb]{0.5,0.5,0.5}}%
    \else
      \def\colorrgb#1{\color{black}}%
      \def\colorgray#1{\color[gray]{#1}}%
      \expandafter\def\csname LTw\endcsname{\color{white}}%
      \expandafter\def\csname LTb\endcsname{\color{black}}%
      \expandafter\def\csname LTa\endcsname{\color{black}}%
      \expandafter\def\csname LT0\endcsname{\color{black}}%
      \expandafter\def\csname LT1\endcsname{\color{black}}%
      \expandafter\def\csname LT2\endcsname{\color{black}}%
      \expandafter\def\csname LT3\endcsname{\color{black}}%
      \expandafter\def\csname LT4\endcsname{\color{black}}%
      \expandafter\def\csname LT5\endcsname{\color{black}}%
      \expandafter\def\csname LT6\endcsname{\color{black}}%
      \expandafter\def\csname LT7\endcsname{\color{black}}%
      \expandafter\def\csname LT8\endcsname{\color{black}}%
    \fi
  \fi
    \setlength{\unitlength}{0.0500bp}%
    \ifx\gptboxheight\undefined%
      \newlength{\gptboxheight}%
      \newlength{\gptboxwidth}%
      \newsavebox{\gptboxtext}%
    \fi%
    \setlength{\fboxrule}{0.5pt}%
    \setlength{\fboxsep}{1pt}%
\begin{picture}(7200.00,4032.00)%
    \gplgaddtomacro\gplbacktext{%
      \csname LTb\endcsname
      \put(588,504){\makebox(0,0)[r]{\strut{}$-30$}}%
      \put(588,1055){\makebox(0,0)[r]{\strut{}$-25$}}%
      \put(588,1606){\makebox(0,0)[r]{\strut{}$-20$}}%
      \put(588,2158){\makebox(0,0)[r]{\strut{}$-15$}}%
      \put(588,2709){\makebox(0,0)[r]{\strut{}$-10$}}%
      \put(588,3260){\makebox(0,0)[r]{\strut{}$-5$}}%
      \put(588,3811){\makebox(0,0)[r]{\strut{}$0$}}%
      \put(1592,284){\makebox(0,0){\strut{}$0.8$}}%
      \put(2655,284){\makebox(0,0){\strut{}$1$}}%
      \put(3719,284){\makebox(0,0){\strut{}$1.2$}}%
      \put(4782,284){\makebox(0,0){\strut{}$1.4$}}%
      \put(5846,284){\makebox(0,0){\strut{}$1.6$}}%
    }%
    \gplgaddtomacro\gplfronttext{%
      \csname LTb\endcsname
      \put(2436,3257){\makebox(0,0){\strut{}$\delta^{3/2}_0(s)$}}%
      \put(3761,64){\makebox(0,0){\strut{}$\sqrt{s}$ GeV}}%
      \csname LTb\endcsname
      \put(5816,3638){\makebox(0,0)[r]{\strut{}UFD}}%
      \csname LTb\endcsname
      \put(5816,3418){\makebox(0,0)[r]{\strut{}CFD}}%
      \csname LTb\endcsname
      \put(5816,3198){\makebox(0,0)[r]{\strut{}Estabrooks et al.}}%
      \csname LTb\endcsname
      \put(5816,2978){\makebox(0,0)[r]{\strut{}Jongejans et al.}}%
      \csname LTb\endcsname
      \put(5816,2758){\makebox(0,0)[r]{\strut{}Cho et al.}}%
      \csname LTb\endcsname
      \put(5816,2538){\makebox(0,0)[r]{\strut{}Linglin et al.}}%
      \csname LTb\endcsname
      \put(5816,2318){\makebox(0,0)[r]{\strut{}Bakker et al.}}%
    }%
    \gplbacktext
    \put(0,0){\includegraphics{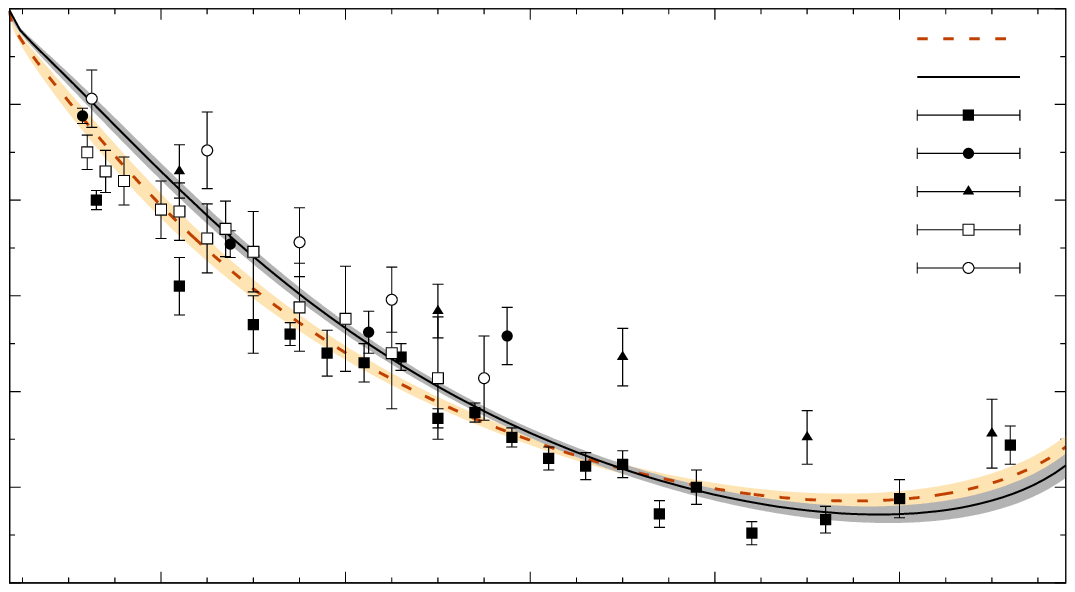}}%
    \gplfronttext
  \end{picture}%
\endgroup

%% file: figures/selascfd.tex
\begingroup
  \makeatletter
  \providecommand\color[2][]{%
    \GenericError{(gnuplot) \space\space\space\@spaces}{%
      Package color not loaded in conjunction with
      terminal option `colourtext'%
    }{See the gnuplot documentation for explanation.%
    }{Either use 'blacktext' in gnuplot or load the package
      color.sty in LaTeX.}%
    \renewcommand\color[2][]{}%
  }%
  \providecommand\includegraphics[2][]{%
    \GenericError{(gnuplot) \space\space\space\@spaces}{%
      Package graphicx or graphics not loaded%
    }{See the gnuplot documentation for explanation.%
    }{The gnuplot epslatex terminal needs graphicx.sty or graphics.sty.}%
    \renewcommand\includegraphics[2][]{}%
  }%
  \providecommand\rotatebox[2]{#2}%
  \@ifundefined{ifGPcolor}{%
    \newif\ifGPcolor
    \GPcolorfalse
  }{}%
  \@ifundefined{ifGPblacktext}{%
    \newif\ifGPblacktext
    \GPblacktexttrue
  }{}%
  \let\gplgaddtomacro\g@addto@macro
  \gdef\gplbacktext{}%
  \gdef\gplfronttext{}%
  \makeatother
  \ifGPblacktext
    \def\colorrgb#1{}%
    \def\colorgray#1{}%
  \else
    \ifGPcolor
      \def\colorrgb#1{\color[rgb]{#1}}%
      \def\colorgray#1{\color[gray]{#1}}%
      \expandafter\def\csname LTw\endcsname{\color{white}}%
      \expandafter\def\csname LTb\endcsname{\color{black}}%
      \expandafter\def\csname LTa\endcsname{\color{black}}%
      \expandafter\def\csname LT0\endcsname{\color[rgb]{1,0,0}}%
      \expandafter\def\csname LT1\endcsname{\color[rgb]{0,1,0}}%
      \expandafter\def\csname LT2\endcsname{\color[rgb]{0,0,1}}%
      \expandafter\def\csname LT3\endcsname{\color[rgb]{1,0,1}}%
      \expandafter\def\csname LT4\endcsname{\color[rgb]{0,1,1}}%
      \expandafter\def\csname LT5\endcsname{\color[rgb]{1,1,0}}%
      \expandafter\def\csname LT6\endcsname{\color[rgb]{0,0,0}}%
      \expandafter\def\csname LT7\endcsname{\color[rgb]{1,0.3,0}}%
      \expandafter\def\csname LT8\endcsname{\color[rgb]{0.5,0.5,0.5}}%
    \else
      \def\colorrgb#1{\color{black}}%
      \def\colorgray#1{\color[gray]{#1}}%
      \expandafter\def\csname LTw\endcsname{\color{white}}%
      \expandafter\def\csname LTb\endcsname{\color{black}}%
      \expandafter\def\csname LTa\endcsname{\color{black}}%
      \expandafter\def\csname LT0\endcsname{\color{black}}%
      \expandafter\def\csname LT1\endcsname{\color{black}}%
      \expandafter\def\csname LT2\endcsname{\color{black}}%
      \expandafter\def\csname LT3\endcsname{\color{black}}%
      \expandafter\def\csname LT4\endcsname{\color{black}}%
      \expandafter\def\csname LT5\endcsname{\color{black}}%
      \expandafter\def\csname LT6\endcsname{\color{black}}%
      \expandafter\def\csname LT7\endcsname{\color{black}}%
      \expandafter\def\csname LT8\endcsname{\color{black}}%
    \fi
  \fi
    \setlength{\unitlength}{0.0500bp}%
    \ifx\gptboxheight\undefined%
      \newlength{\gptboxheight}%
      \newlength{\gptboxwidth}%
      \newsavebox{\gptboxtext}%
    \fi%
    \setlength{\fboxrule}{0.5pt}%
    \setlength{\fboxsep}{1pt}%
\begin{picture}(7200.00,4032.00)%
    \gplgaddtomacro\gplbacktext{%
      \csname LTb\endcsname
      \put(588,504){\makebox(0,0)[r]{\strut{}$0$}}%
      \put(588,1105){\makebox(0,0)[r]{\strut{}$10$}}%
      \put(588,1707){\makebox(0,0)[r]{\strut{}$20$}}%
      \put(588,2308){\makebox(0,0)[r]{\strut{}$30$}}%
      \put(588,2909){\makebox(0,0)[r]{\strut{}$40$}}%
      \put(588,3510){\makebox(0,0)[r]{\strut{}$50$}}%
      \put(959,284){\makebox(0,0){\strut{}$0.65$}}%
      \put(1811,284){\makebox(0,0){\strut{}$0.7$}}%
      \put(2662,284){\makebox(0,0){\strut{}$0.75$}}%
      \put(3514,284){\makebox(0,0){\strut{}$0.8$}}%
      \put(4366,284){\makebox(0,0){\strut{}$0.85$}}%
      \put(5218,284){\makebox(0,0){\strut{}$0.9$}}%
      \put(6070,284){\makebox(0,0){\strut{}$0.95$}}%
    }%
    \gplgaddtomacro\gplfronttext{%
      \csname LTb\endcsname
      \put(1908,2157){\makebox(0,0){$\delta^{1/2}_0(s)$}}%
      \put(3761,64){\makebox(0,0){$\sqrt{s}$ GeV}}%
      \csname LTb\endcsname
      \put(3228,3638){\makebox(0,0)[r]{\strut{} UFD}}%
      \csname LTb\endcsname
      \put(3228,3418){\makebox(0,0)[r]{\strut{} CFD}}%
      \csname LTb\endcsname
      \put(3228,3198){\makebox(0,0)[r]{\strut{} Estabrooks et al.}}%
      \csname LTb\endcsname
      \put(3228,2978){\makebox(0,0)[r]{\strut{} Aston et al.}}%
    }%
    \gplbacktext
    \put(0,0){\includegraphics{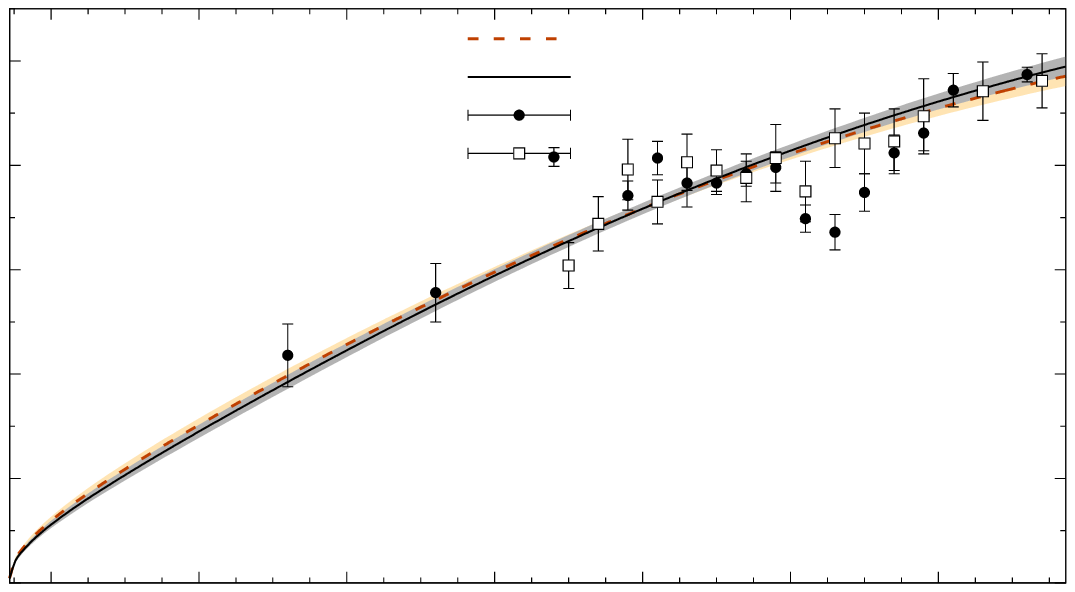}}%
    \gplfronttext
  \end{picture}%
\endgroup

%% file: figures/sclcfd.tex
\begingroup
  \makeatletter
  \providecommand\color[2][]{%
    \GenericError{(gnuplot) \space\space\space\@spaces}{%
      Package color not loaded in conjunction with
      terminal option `colourtext'%
    }{See the gnuplot documentation for explanation.%
    }{Either use 'blacktext' in gnuplot or load the package
      color.sty in LaTeX.}%
    \renewcommand\color[2][]{}%
  }%
  \providecommand\includegraphics[2][]{%
    \GenericError{(gnuplot) \space\space\space\@spaces}{%
      Package graphicx or graphics not loaded%
    }{See the gnuplot documentation for explanation.%
    }{The gnuplot epslatex terminal needs graphicx.sty or graphics.sty.}%
    \renewcommand\includegraphics[2][]{}%
  }%
  \providecommand\rotatebox[2]{#2}%
  \@ifundefined{ifGPcolor}{%
    \newif\ifGPcolor
    \GPcolortrue
  }{}%
  \@ifundefined{ifGPblacktext}{%
    \newif\ifGPblacktext
    \GPblacktexttrue
  }{}%
  \let\gplgaddtomacro\g@addto@macro
  \gdef\gplbacktext{}%
  \gdef\gplfronttext{}%
  \makeatother
  \ifGPblacktext
    \def\colorrgb#1{}%
    \def\colorgray#1{}%
  \else
    \ifGPcolor
      \def\colorrgb#1{\color[rgb]{#1}}%
      \def\colorgray#1{\color[gray]{#1}}%
      \expandafter\def\csname LTw\endcsname{\color{white}}%
      \expandafter\def\csname LTb\endcsname{\color{black}}%
      \expandafter\def\csname LTa\endcsname{\color{black}}%
      \expandafter\def\csname LT0\endcsname{\color[rgb]{1,0,0}}%
      \expandafter\def\csname LT1\endcsname{\color[rgb]{0,1,0}}%
      \expandafter\def\csname LT2\endcsname{\color[rgb]{0,0,1}}%
      \expandafter\def\csname LT3\endcsname{\color[rgb]{1,0,1}}%
      \expandafter\def\csname LT4\endcsname{\color[rgb]{0,1,1}}%
      \expandafter\def\csname LT5\endcsname{\color[rgb]{1,1,0}}%
      \expandafter\def\csname LT6\endcsname{\color[rgb]{0,0,0}}%
      \expandafter\def\csname LT7\endcsname{\color[rgb]{1,0.3,0}}%
      \expandafter\def\csname LT8\endcsname{\color[rgb]{0.5,0.5,0.5}}%
    \else
      \def\colorrgb#1{\color{black}}%
      \def\colorgray#1{\color[gray]{#1}}%
      \expandafter\def\csname LTw\endcsname{\color{white}}%
      \expandafter\def\csname LTb\endcsname{\color{black}}%
      \expandafter\def\csname LTa\endcsname{\color{black}}%
      \expandafter\def\csname LT0\endcsname{\color{black}}%
      \expandafter\def\csname LT1\endcsname{\color{black}}%
      \expandafter\def\csname LT2\endcsname{\color{black}}%
      \expandafter\def\csname LT3\endcsname{\color{black}}%
      \expandafter\def\csname LT4\endcsname{\color{black}}%
      \expandafter\def\csname LT5\endcsname{\color{black}}%
      \expandafter\def\csname LT6\endcsname{\color{black}}%
      \expandafter\def\csname LT7\endcsname{\color{black}}%
      \expandafter\def\csname LT8\endcsname{\color{black}}%
    \fi
  \fi
    \setlength{\unitlength}{0.0500bp}%
    \ifx\gptboxheight\undefined%
      \newlength{\gptboxheight}%
      \newlength{\gptboxwidth}%
      \newsavebox{\gptboxtext}%
    \fi%
    \setlength{\fboxrule}{0.5pt}%
    \setlength{\fboxsep}{1pt}%
\begin{picture}(12440.00,5640.00)%
    \gplgaddtomacro\gplbacktext{%
      \csname LTb\endcsname
      \put(1754,1205){\makebox(0,0)[r]{\strut{}$-0.07$}}%
      \csname LTb\endcsname
      \put(1754,2262){\makebox(0,0)[r]{\strut{}$-0.06$}}%
      \csname LTb\endcsname
      \put(1754,3320){\makebox(0,0)[r]{\strut{}$-0.05$}}%
      \csname LTb\endcsname
      \put(1754,4377){\makebox(0,0)[r]{\strut{}$-0.04$}}%
      \csname LTb\endcsname
      \put(1754,5435){\makebox(0,0)[r]{\strut{}$-0.03$}}%
      \csname LTb\endcsname
      \put(1866,472){\makebox(0,0){\strut{}$0.12$}}%
      \csname LTb\endcsname
      \put(3021,472){\makebox(0,0){\strut{}$0.14$}}%
      \csname LTb\endcsname
      \put(4176,472){\makebox(0,0){\strut{}$0.16$}}%
      \csname LTb\endcsname
      \put(5331,472){\makebox(0,0){\strut{}$0.18$}}%
      \csname LTb\endcsname
      \put(6486,472){\makebox(0,0){\strut{}$0.2$}}%
      \csname LTb\endcsname
      \put(7641,472){\makebox(0,0){\strut{}$0.22$}}%
      \csname LTb\endcsname
      \put(8796,472){\makebox(0,0){\strut{}$0.24$}}%
      \csname LTb\endcsname
      \put(9951,472){\makebox(0,0){\strut{}$0.26$}}%
    }%
    \gplgaddtomacro\gplfronttext{%
      \csname LTb\endcsname
      \put(1008,3055){\rotatebox{-270}{\makebox(0,0){\strut{}$m_{\pi} a^{3/2}_0$}}}%
      \csname LTb\endcsname
      \put(5908,268){\makebox(0,0){\strut{}$m_{\pi} a^{1/2}_0$}}%
      \csname LTb\endcsname
      \put(11432,5189){\makebox(0,0)[r]{\strut{}\small{ Miao et al. 04}}}%
      \csname LTb\endcsname
      \put(11432,4966){\makebox(0,0)[r]{\strut{}\footnotesize DIRAC 17}}%
      \csname LTb\endcsname
      \put(11432,4743){\makebox(0,0)[r]{\strut{}\footnotesize HDR SR CFD}}%
      \csname LTb\endcsname
      \put(11432,4520){\makebox(0,0)[r]{\strut{}\footnotesize FDR SR CFD}}%
      \csname LTb\endcsname
      \put(11432,4297){\makebox(0,0)[r]{\strut{}\footnotesize FN 07}}%
      \csname LTb\endcsname
      \put(11432,4074){\makebox(0,0)[r]{\strut{}\footnotesize Buettiker et al. 04}}%
      \csname LTb\endcsname
      \put(11432,3851){\makebox(0,0)[r]{\strut{}\footnotesize BE LO 14}}%
      \csname LTb\endcsname
      \put(11432,3628){\makebox(0,0)[r]{\strut{}\footnotesize BE NLO 14}}%
      \csname LTb\endcsname
      \put(11432,3405){\makebox(0,0)[r]{\strut{}\footnotesize BE NLO ff 14}}%
      \csname LTb\endcsname
      \put(11432,3182){\makebox(0,0)[r]{\strut{}\footnotesize BE NNLO 14}}%
      \csname LTb\endcsname
      \put(11432,2959){\makebox(0,0)[r]{\strut{}\footnotesize BE NNLO ff 14}}%
      \csname LTb\endcsname
      \put(11432,2736){\makebox(0,0)[r]{\strut{}\footnotesize NPLQCD 06}}%
      \csname LTb\endcsname
      \put(11432,2513){\makebox(0,0)[r]{\strut{}\footnotesize FU 12}}%
      \csname LTb\endcsname
      \put(11432,2290){\makebox(0,0)[r]{\strut{}\footnotesize PACS-CS 14}}%
      \csname LTb\endcsname
      \put(11432,2067){\makebox(0,0)[r]{\strut{}\footnotesize ETM 18}}%
      \csname LTb\endcsname
      \put(11432,1844){\makebox(0,0)[r]{\strut{}\footnotesize FDR CFD-old}}%
      \csname LTb\endcsname
      \put(11432,1621){\makebox(0,0)[r]{\strut{}\footnotesize  FINAL VALUE}}%
      \csname LTb\endcsname
      \put(11432,1398){\makebox(0,0)[r]{\strut{}\footnotesize New CFD}}%
    }%
    \gplbacktext
    \put(0,0){\includegraphics{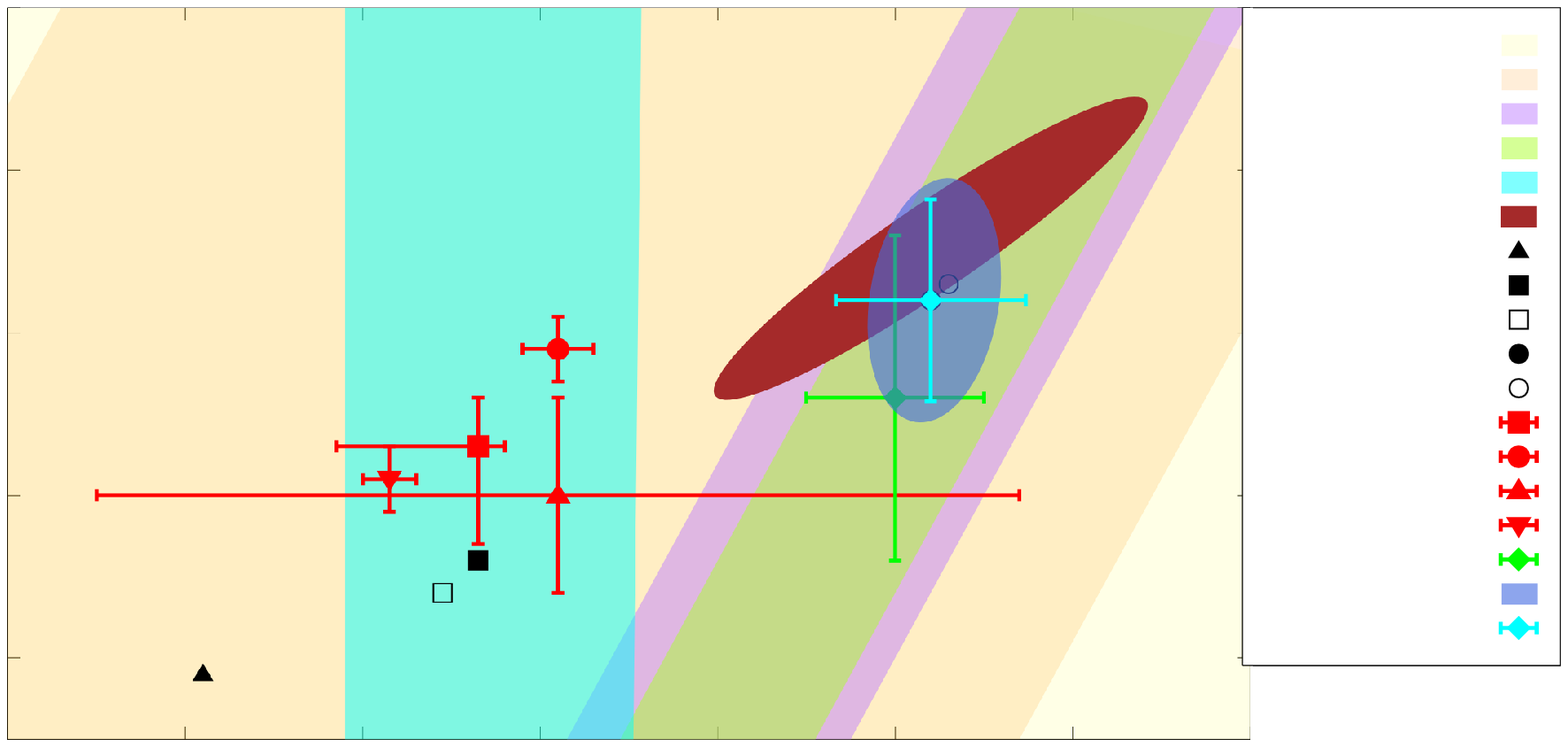}}%
    \gplfronttext
  \end{picture}%
\endgroup